\documentclass[twocolumn]{aastex62}
\hypersetup{linkcolor=blue,citecolor=blue,filecolor=cyan,urlcolor=blue}

\usepackage{longtable}
\usepackage{lipsum} 
\usepackage{multirow}
\usepackage[caption=false]{subfig}
\graphicspath{{./}{figures/}}

\def\arcsec{$^{\prime\prime}$}
\def\um{$\mu \rm m$}
\def\Mo{$M_{\odot}$}
\newcommand{\HII}{H\textsc{ii}}

\newcommand\new[1]{\textcolor{black}{\textbf{#1}}}

\begin{document}

\title{ALMA observations of NGC 6334S $-$ I: \\
Forming massive stars and cluster in subsonic and transonic filamentary clouds}

\correspondingauthor{Shanghuo Li}
\email{shanghuo.li@gmail.com, jzwang@shao.ac.cn}

\author[0000-0003-1275-5251]{Shanghuo Li}
\affil{Shanghai Astronomical Observatory, Chinese Academy of Sciences, 80 Nandan Road, Shanghai 200030, China}
\affiliation{Center for Astrophysics $|$ Harvard \& Smithsonian , 60 Garden Street, Cambridge, MA 02138, USA}
\affiliation{University of Chinese Academy of Sciences, 19A Yuquanlu, Beijing 100049, China}

\author{Qizhou Zhang}
\affiliation{Center for Astrophysics $|$ Harvard \& Smithsonian , 60 Garden Street, Cambridge, MA 02138, USA}

 \author{Hauyu Baobab Liu}
 \affiliation{Academia Sinica Institute of Astronomy and Astrophysics, P.O. Box 23-141, Taipei 106, Taiwan}

 \author{Henrik Beuther}
 \affiliation{Max Planck Institute for Astronomy, Konigstuhl 17, 69117 Heidelberg, Germany}

 \author{Aina Palau}
 \affiliation{Instituto de Radioastronom\'ia y Astrof\'isica, Universidad Nacional Aut\'onoma de M\'exico, P.O. Box 3-72, 58090, Morelia, Michoac\'an, M\'exico}
 
 \author{Josep Miquel. Girart}
 \affiliation{Institut de Ci\`encies de l'Espai (IEEC-CSIC), Campus UAB, Carrer de Can Magrans s/n, 08193 Cerdanyola del Vall\`es, Catalonia, Spain} 

 \author{Howard Smith}
 \affiliation{Center for Astrophysics $|$ Harvard \& Smithsonian , 60 Garden Street, Cambridge, MA 02138, USA}

 \author{Joseph L. Hora}
 \affiliation{Center for Astrophysics $|$ Harvard \& Smithsonian , 60 Garden Street, Cambridge, MA 02138, USA}
 
 \author{Yuxing Lin}
 \affiliation{Max-Planck Institute f\"ur Radio Astronomy, Auf dem H\"ugel 69, 53121, Bonn, Germany}
 
 \author{Keping Qiu}
 \affiliation{School of Astronomy and Space Science, Nanjing University, 163 Xianlin Avenue, Nanjing 210023, China}
 
 \author{Shaye Strom}
 \affiliation{Center for Astrophysics $|$ Harvard \& Smithsonian , 60 Garden Street, Cambridge, MA 02138, USA}

 \author{Junzhi Wang}
 \affil{Shanghai Astronomical Observatory, Chinese Academy of Sciences, 80 Nandan Road, Shanghai 200030, China}

 \author{Fei Li}
\affil{Shanghai Astronomical Observatory, Chinese Academy of Sciences, 80 Nandan Road, Shanghai 200030, China}
\affiliation{University of Chinese Academy of Sciences, 19A Yuquanlu, Beijing 100049, China}
 
 \author{Nannan Yue}
 \affiliation{National Astronomical Observatories, Chinese Academy of Sciences, Beijing 100012}
\affiliation{University of Chinese Academy of Sciences, 19A Yuquanlu, Beijing 100049, China}

\begin{abstract}
We present Atacama Large Millimeter/submillimeter Array 
(ALMA)  and  Karl G. Jansky Very Large Array (JVLA) 
observations of the  massive infrared dark cloud NGC 6334S 
(also known as IRDC G350.56+0.44), 
located at the southwestern 
end of the NGC 6334 molecular cloud  complex.  
The H$^{13}$CO$^{+}$ and the NH$_{2}$D lines covered by 
the ALMA observations at a $\sim$3\arcsec\ angular resolution 
($\sim$0.02 pc) reveal that the spatially unresolved 
non-thermal motions are predominantly subsonic and transonic, a condition  
analogous to that found in low-mass star-forming molecular clouds. 
The observed supersonic non-thermal velocity dispersions in 
massive star forming regions, often reported in the  
literature, might be significantly biased by poor spatial resolutions
that broaden the observed line widths due to unresolved motions 
within the telescope beam. 
Our 3~mm continuum image resolves 49 dense cores, 
whose masses range from 0.17 to 14 \Mo.  The majority of 
them are resolved with multiple velocity components. Our 
analyses of these gas velocity components find an 
anti-correlation between the gas mass and the virial
parameter.
This implies that the more massive structures tend to be 
more gravitationally unstable. 
Finally, we find that the external pressure in the 
NGC 6334S cloud is important in confining these dense structures, and 
 may play a  role in the formation of dense 
cores, and subsequently, the embedded young stars.

\end{abstract}



\section{Introduction} 
\label{sec:intro}
Molecular clouds in the Milky Way are in general inefficient 
in forming stars 
\citep[$\sim$1\%,][]{1986ApJ...301..398M,2011ApJ...729..133M,
2016ApJ...831...73V}, an observation that  has motivated the 
long-standing hypotheses that they are supported by 
supersonic turbulence or magnetic fields against their 
self-gravitational collapse 
\citep[for the review see][]{2000prpl.conf....3V,2004RvMP...76..125M,
2007ARA&A..45..339B,2007ARA&A..45..565M}. 
\citet{1981MNRAS.194..809L} reported an empirical, positive 
correlation between the physical size scale and the 
velocity dispersion of molecular clouds. 
Although the study was based on measurements from different 
clouds, this correlation has been taken as indications that 
turbulence dissipates toward smaller spatial scales where 
gas is prone to the gravitational 
collapse and the subsequent star-formation 
\citep[see][and references therein]{1987A&A...172..293B,
1998ApJ...508L..99S,1998PhRvL..80.2754M,
2004RvMP...76..125M,2005ApJ...630..250K,
2004ARA&A..42..211E,2008ApJ...684..395H}. 
As a consequence, on spatial scales of dense star-forming 
cores ($\lesssim$0.1 pc), the non-thermal motions may be 
sonic \citep{1988ApJ...329..392M,1998ApJ...504..223G,
2002ApJ...572..238C}. 
This is consistent with the observations of some 
filamentary, low-mass star-forming clouds 
\citep[e.g.,][]{2011A&A...533A..34H,
2013A&A...554A..55H,2015Natur.518..213P,
2016A&A...587A..97H,2017A&A...606A.123H}. 

In contrast, previous observations towards high-mass 
star-forming regions have reported supersonic non-thermal  
velocity dispersion 
\citep{1995ApJ...446..665C,2003A&A...405..639P, 
2003ApJS..149..375S,2008ApJ...672L..33W,2011A&A...527A..88V,
2012ApJ...756...60S,2017MNRAS.466..248L}. 
The formation of gas cores of  high mass and high density may 
be supported by  supersonic turbulence. 
Otherwise they may fragment to form lower-mass stars since 
their masses are much larger than the Jeans mass 
\citep{2003ApJ...585..850M}.  However, recent higher angular 
resolution observations started to reveal subsonic non-thermal 
velocity dispersion in massive star-forming molecular clouds 
on small spatial scales  
\citep[e.g.,][]{2018A&A...614A..64B,2018A&A...610A..77H,
2018ApJ...861...77M,2018A&A...611L...3S}.  
Whether the massive star- and 
cluster-forming regions are  initially supported by 
supersonic turbulence remains  a matter of debate, and needs 
to be clarified with more observations of target sources at 
early evolutionary stages.

\begin{figure*}[!ht]
\centering
\includegraphics[angle=90,scale=0.38]{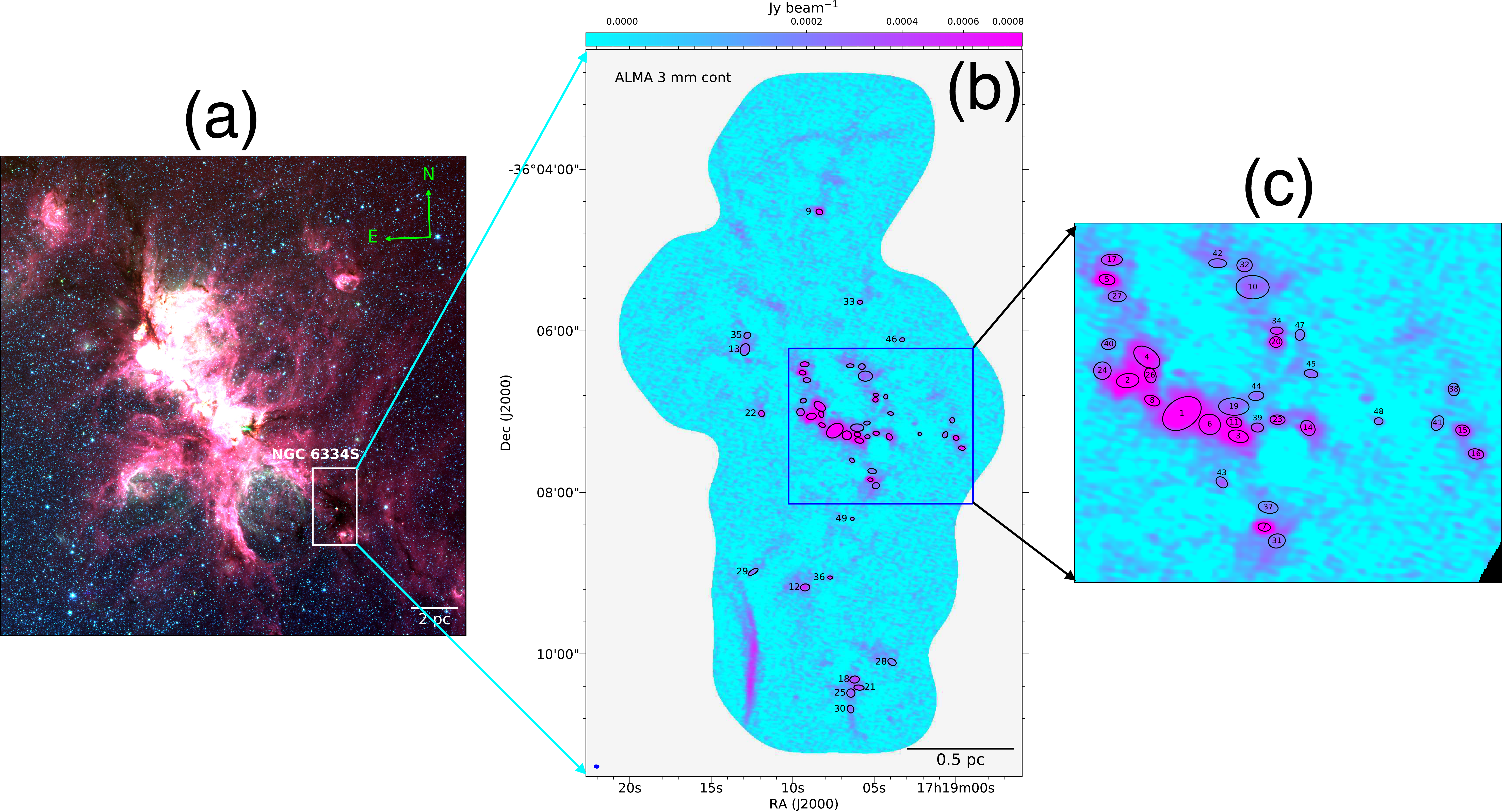}
\caption{
Panel a: three-color \texttt{Spitzer} composite image of the 
NGC 6334 molecular cloud complex with blue, green, and red 
representing data at $\lambda$ = 3.6, 4.5, and 8.0 \um, respectively. 
The image is centered at 17h20m02s, -35d54m03s. The scale bar is 
approximately 2 pc at the distance of 1.3 kpc. The white box presents 
the NGC 6334S region. 
Panel b: the ALMA 3~mm continuum image. The beam size is 
shown in the bottom left of the panel. 
Panel c: zoom in of the central region.  
The black ellipses present the identified dense cores. 
}
 \label{rgb}
\end{figure*}

Using the Atacama Large Millimeter Array (ALMA) and the 
Karl G. Jansky Very Large Array (JVLA),
we have carried out  high angular resolution observations 
towards the infrared dark, massive cluster-forming molecular 
cloud NGC 6334S, which is located at the southwestern end of the 
NGC\,6334 molecular cloud complex.  
In contrast to the other luminous OB 
cluster-forming clumps in the NGC\,6334 molecular cloud 
complex, namely the I, I(N), II, III, IV, and V clumps 
\citep{2008hsf2.book..456P,2013A&A...554A..42R,2013ApJ...778...96W}, 
NGC 6334S is dark 
at IR wavelengths and lacks signs of massive star 
formation (see Figure \ref{rgb}). With a mass of 
1.3 $\times$ 10$^{3}$ \Mo, comparable to the clumps with 
embedded massive protostars and protocluster in the complex, 
NGC~6334S has the potential to form massive stars with a cluster 
of lower mass objects. The proximity of NGC~6334S 
\citep[$d\sim$1.3 kpc;][]{2014ApJ...784..114C} 
makes it an ideal laboratory to investigate the key physical 
processes related to massive star and cluster formation.

In this paper, we investigate the dynamical motions of 
identified embedded dense cores and their parent cloud, 
as well as the dynamical stability of dense cores. 
First, we describe our ALMA and JVLA observations in $\S$ 
\ref{sec:obs}. Then, we present the results and analysis 
in $\S$ \ref{results}. We discuss in detail the properties 
of dense cores and their parent cloud in $\S$ \ref{discussion}.  
Finally,  we summarize the conclusion 
of this work in $\S$ \ref{conclusion}.

\section{Observations} 
\label{sec:obs}
\subsection{ALMA Observations} 
We carried out a 55-pointings mosaic  of 
the massive infrared dark cloud (IRDC), NGC 6334S, between 2017 
March 13 and 2017 March 21 using the 12-m main array of ALMA 
(ID: 2016.1.00951.S, PI: Shaye Strom). 
The overall observing time and the on-source integration time 
are 7.3 hours and 4.4 hours, respectively.  The  projected 
baseline lengths range from 15 to 155 meters 
($\sim$4.9-51 k$\lambda$ at the averaged frequency of 98.5 GHz).

We employed two 234.4 MHz wide spectral windows to cover the 
H$^{13}$CO$^{+}$ 1-0 (86.754 GHz) and 
NH$_{2}$D $1_{11}$-$1_{01}$ (85.926 GHz) 
lines, respectively, with a spectral resolution of 61 kHz 
($\sim$0.21 km\,s$^{-1}$ at 86 GHz).  In addition, we 
centered three 1.875 GHz wide spectral windows at 88.5 GHz, 
98.5 GHz, and 100.3 GHz to obtain broad band continuum data. 
We observed the quasars J1617-5858, J1713-3418, and 
J1733-1304, for passband, complex gain, and absolute flux 
calibrations, respectively.

The data calibrations were performed by the supporting 
staff at the ALMA Regional Center (ARC), using the CASA 
software package \citep{2007ASPC..376..127M}. 
The continuum image was obtained using the three 
1.875~GHz spectral windows with a Briggs's robust weighting 
of 0.5 to the visibilities. This yields a synthesized beam of 
3\arcsec.6 $\times$ 2\arcsec.4 
(or 0.023 $\times$ 0.015 pc)  
with a position angle (P.A.) of 
81$^{\circ}$ and a 1$\sigma$ root mean square (rms) noise 
level of 0.03 mJy\,beam$^{-1}$. 
For both H$^{13}$CO$^{+}$ and NH$_{2}$D lines, we used a 
Briggs's robust weighting of 0.5 to the visibilities, which achieved 
a synthesized beam of about 4\arcsec.1 $\times$ 2\arcsec.8 
(PA= 83$^{\circ}$, 
or  0.026 $\times$ 0.018 pc) 
for these 
two lines. The 1$\sigma$ rms noise level is about 
6 mJy\,beam$^{-1}$ per 0.21 km\,s$^{-1}$ channel for both the 
H$^{13}$CO$^{+}$ and NH$_{2}$D lines.

\begin{figure*}[ht!]
\epsscale{1.2}
\plotone{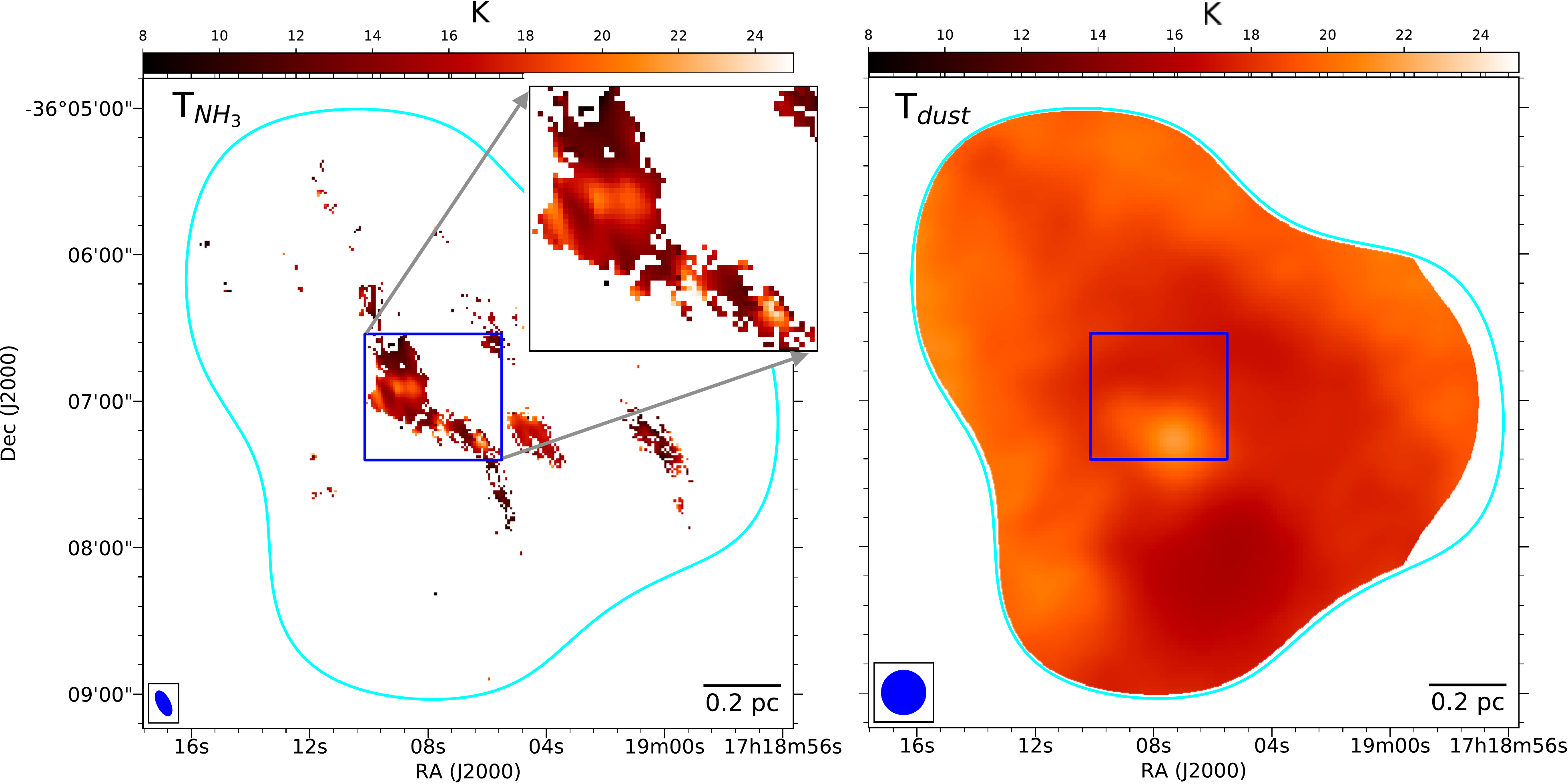}
\caption{
Left:  temperatures derived from the  NH$_{3}$ data. The cyan line is
the JVLA mosaic pattern. 
Right: the dust temperatures derived from SED fitting. 
The blue box marks the zooming in region in the left panel. 
The beam size is shown in the bottom left of the panel.
\label{temperature}}
\end{figure*}

\subsection{JVLA Observations} 
We carried out a 4-pointings mosaic of the 
central region of NGC 6334S using the JVLA, on 2014 August 28 
(project code: 14A-241; PI: Qizhou Zhang). 
These observations simultaneously covered the NH$_{3}$ (1,1) 
through (5,5) metastable inversion transitions.  The overall 
observing time was 2 hours, which yielded an on-source 
integration of $\sim$20 minutes for each pointing.  The Quasars 
3C286, J1743-0350 and J1744-3316 were observed for flux,  
bandpass and gain calibrations, respectively.

The data calibration and imaging were carried out using the 
CASA software package.  Our target source is located in the 
south, which leads to shorter projected baselines in the 
north-south direction. To increase the signal-to-noise (S/N) 
ratio, we tapered the visibilities with an elliptical 
Gaussian with full width at half-maximum (FWHM) of  
6\arcsec $\times$ 3\arcsec\ (P.A. = 0$^{\circ}$).
The achieved synthesized beam size is about 10\arcsec $\times$ 
5\arcsec\ (P.A. = 26$^{\circ}$, 
 or  0.063 $\times$ 0.032 pc) 
for the NH$_{3}$ (1,1) and (2,2) lines. 
The continuum image achieved a 1$\sigma$ rms noise of 
30 $\mu$Jy\,beam$^{-1}$, while spectral line images achieved a 
1$\sigma$ rms noise of 9 mJy\,beam$^{-1}$ at a spectral 
resolution  of  0.2 km\,s$^{-1}$.

\subsection{Fitting molecular line spectra}
\label{obs:fitting}
We fit Gaussian line profiles to the  H$^{13}$CO$^{+}$ image 
cube pixel by pixel, using the \texttt{PySpecKit} package 
\citep{2011ascl.soft09001G}.  We excluded regions where the 
peak intensity in the spectral domain is below 5 times the 
rms noise level. For data above the cutoff, we regarded 
spectral peaks that are separated by more than 2 spectral 
channels  and brighter than 5 times the rms noise level as 
independent velocity components.  We then simultaneously fit 
Gaussian profiles to all of the independent velocity 
components. We eliminated poor fits by only keeping pixels 
that fulfill the following criteria:  
(1) $\sigma_{\rm obs}$ $>$ 3$\Delta_{\sigma_{\rm obs}}$, 
which ensures that the derived  observed velocity 
dispersion is reliable.
(2) -8 $\leqslant$ $v_{\rm lsr} \leqslant$ 0 km s$^{-1}$, 
which is the H$^{13}$CO$^{+}$ line emission velocity range 
in  the NGC 6334S. 
(3) $I >$ 3$\Delta_{I}$. 
(In these criteria, $\Delta_{\sigma_{\rm obs}}$ is the 
uncertainty of the  observed velocity dispersion 
$\sigma_{\rm obs}$, $v_{\rm lsr}$ is the central velocity, 
and $\Delta_{I}$ is the uncertainty of velocity 
integrated intensity $I$.) The  observed velocity dispersion, 
$\sigma_{\rm obs}$, is the combination of both thermal and 
non-thermal velocity dispersion values. 
The NH$_{2}$D line cubes were fit following a similar routine 
but including its 36 hyperfine components 
\citep{2016A&A...586L...4D}.

The NH$_{3}$ (1,1) and (2,2) transitions were jointly fit, 
pixel by pixel, using \texttt{PySpecKit} in order to estimate
the gas kinetic temperature and the line width. 
Details of the procedure have been provided in 
\cite{2017ApJ...843...63F}. 
Limited by S/N ratios, we cannot robustly fit multiple 
velocity components to the NH$_{3}$ spectra. 
Nevertheless, for our purpose of deriving the gas temperature 
and assessing the uncertainty of converting the NH$_{3}$ 
rotational temperature to the gas kinetic temperature, it 
is adequate to assume a single velocity component. 
We have compared the gas temperatures estimated from 
NH$_{3}$ and dust temperatures estimated by SED fitting the 
\texttt{PACS} 160 \um, \texttt{SPIRE} 250 \um, 
\texttt{PLANCK} and ATLASGAL 870 \um\  data 
(see Appendix \ref{SED}).  
In spite of the different angular resolutions, we found that 
the derived gas and dust temperatures are consistent to 
within $\leqslant$ 3 K at the central part of NGC 6334S, and are 
consistent to around 6 K over a more extended area 
(Figure \ref{temperature}).  We use a uniform temperature of 
$\langle T_{\rm NH_{3}} \rangle$ = 15 K, the averaged temperature 
from the NH$_{3}$ data, to calculate the gas masses, thermal line widths 
and the sound speeds for the regions where the NH$_{3}$ data 
are not available (see Section \ref{results:linewidth} and 
Section \ref{results:virial}). 

\begin{figure*}[ht!]
\centering
\includegraphics[angle=0,scale=0.36]{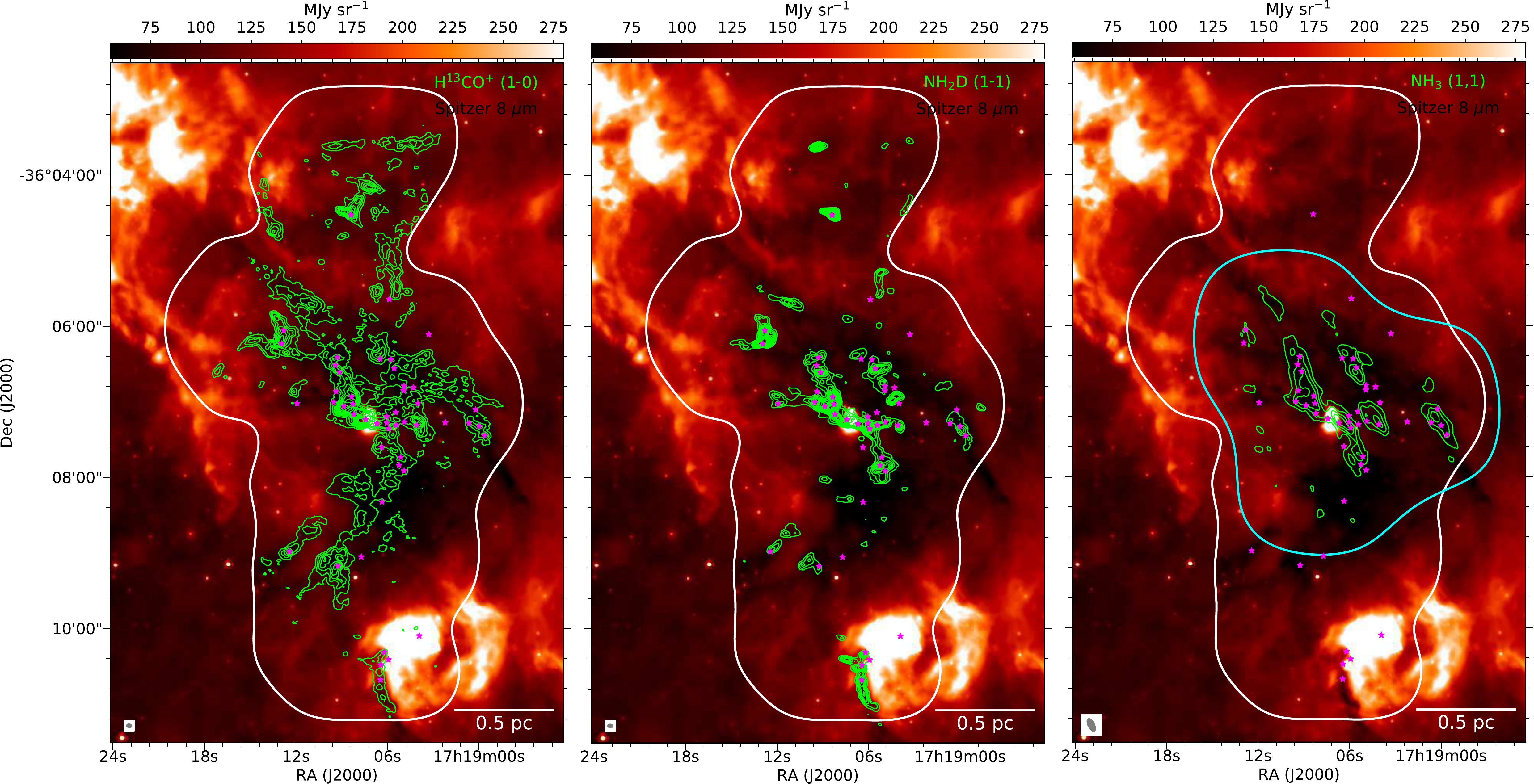}
\caption{
Figure shows the velocity integrated 
intensity map (green contours) of the H$^{13}$CO$^{+}$ (1-0), 
NH$_{2}$D (1-1), and NH$_{3}$ (1,1) overlaid on the Spitzer 8 \um\ 
emission.  The green contours are 
$\pm$(3, 6, 9 ....) $\times$ $\sigma$, where $\sigma$ is
the rms level for each line. The pink stars mark the dense cores 
revealed in the 3~mm continuum data. The white and cyan solid lines 
show the  ALMA and VLA primary-beam responses at 20\%, 
respectively.  The beam size is shown in the bottom right of the panel. 
\label{mom0}}
\end{figure*}

\section{Results and Analysis} 
\label{results}
The present work focuses on the non-thermal velocity dispersion 
of the NGC\,6334S cloud and the embedded dense cores, as 
well as the stability of the dense cores. 
Both H$^{13}$CO$^{+}$ 
1-0 (critical density, n$_{\rm cr} \sim$ 2 $\times $10$^{4}$ 
cm$^{-3}$) and NH$_{2}$D $1_{11}$-$1_{01}$ 
(n$_{\rm cr} \sim$ a few 10$^{5}$ cm$^{-3}$) are good dense gas 
tracers in star formation regions 
\citep[e.g.,][]{2007A&A...467..207P,2010A&A...517L...6B,
2012ApJ...756...60S}, which 
allows us to investigate the kinematic properties on spatial 
scales of both molecular cloud and dense cores. The 
H$^{13}$CO$^{+}$ molecule 
($\mu$ = 30, $\sigma_{\rm th(15\rm K)}$ = 0.06 km s$^{-1}$) 
has a higher molecular weight than the NH$_{2}$D molecule 
($\mu$ = 18, $\sigma_{\rm th(15\rm K)}$ = 0.08 km s$^{-1}$), 
and therefore is a better probe of the non-thermal gas 
motions.  
In spite of the hyperfine line splitting of the 
H$^{13}$CO$^{+}$ 1-0 transitions with six components, 
they are closer than 0.14 km s$^{-1}$ (41 kHz) at the 
frequency of the H$^{13}$CO$^{+}$ 1-0 line 
\citep{2004A&A...419..949S}. 
The averaged line width reduction is about 0.01 km s$^{-1}$ 
when the hyperfine structure of the H$^{13}$CO$^{+}$ is fully 
taken into account, which is much smaller than the derived 
line widths  (Section \ref{results:linewidth}). 
Therefore, the hyperfine components of the 
H$^{13}$CO$^{+}$ 1-0 can be treated as the same transition 
for our analysis. 
The  main hyperfine line components  
of the NH$_{2}$D 1-1 line 
are resolved by the observations, which permits robustly 
deriving the line widths towards the localized cores with  
higher gas column densities \citep[e.g.,][]{2011A&A...530A.118P}. 
When there are multiple velocity components along the 
line-of-sight, they can be confused with the hyperfine line 
structures of the NH$_{2}$D. Nevertheless, this issue can be 
mitigated by comparing them with the spectrum of the 
H$^{13}$CO$^{+}$ line.  Finally, the multiple inversion line 
transitions of NH$_{3}$ can be simultaneously covered by 
the observations of JVLA thanks to its  broad bandwidth 
capability.

\subsection{Dense core identification}
\label{results:identification}
We employed the 
{\tt astrodendro}\footnote{\url{http://dendrograms.org/}} 
algorithm to pre-select dense cores (i.e., the leaves in 
the terminology of  {\tt astrodendro}) 
from the 3~mm continuum image (Figure \ref{rgb}). 
The specific properties of the dense cores (size, flux 
density, peak intensity and position) identified from the 
{\tt astrodendro} analysis are obtained using the 
CASA-{\tt imfit} task.  There are some compact sources that are 
detected at $>$ 5$\sigma$ significance but missed by 
{\tt astrodendro};  we use CASA-{\tt imfit} to recover them 
from the image.  We avoid identifying sources from the 
elongated filament southeast of  NGC 6334S (Figure \ref{rgb}).  
This elongated filament is not detected with spectral line 
emission counterpart, and is likely dominated by  free-free or 
synchrotron emission and likely associated with the 
\HII\ region east of this filament. When computing the 
dendrogram the following parameters are used: the minimum 
pixel value $\mathbf{min_{-}value}$ = 3$\sigma$, where 
$\sigma$ is the rms noise of continuum image;  the minimum 
difference in the peak intensity between neighboring compact 
structures $\mathbf{min_{-}delta}$ = 1$\sigma$;  the 
minimum number of pixels required for a structure to be 
considered an independent entity $\mathbf{min_{-}npix}$ 
= 40, which is approximately the synthesized beam area.

We identified 49 dense cores (Table \ref{tab:cores}), which 
are labelled in Figure \ref{rgb} and presented as pink stars in 
Figure \ref{mom0}.  They are closely associated with 
gas filamentary structures, and the majority (39) of these 
cores are concentrated in the central 1 pc area of the image.

\subsection{Intensity distributions of molecular gas tracers}
\label{results:lines}

Figure \ref{mom0} shows the velocity integrated intensity 
maps of the H$^{13}$CO$^{+}$, NH$_{2}$D and NH$_{3}$ 
emission, overlaid with the \textit{Spitzer} 8 \um\ emission. 
We detected significant H$^{13}$CO$^{+}$ line emission 
coinciding with the 8 \um\ dark filamentary structures, 
as well as with the majority of embedded dense cores. 
On the other hand, the NH$_{2}$D and NH$_{3}$ line emissions 
are preferentially detected at the location of dense cores.

\begin{figure*}[ht!]
\epsscale{1.2}
\plotone{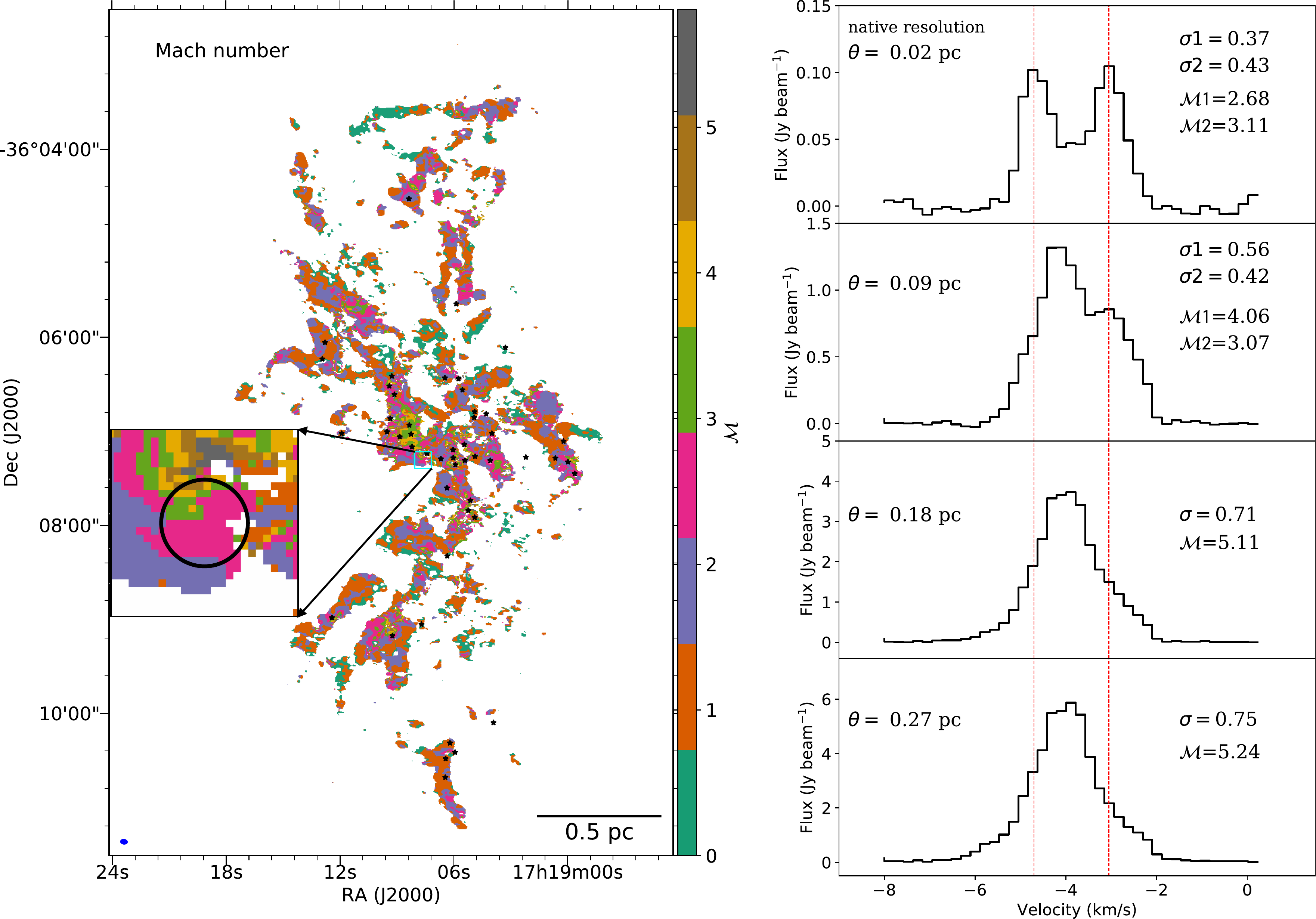}
\caption{
Left: Mach number map estimated from the  
H$^{13}$CO$^{+}$ line for NGC 6334S. The black stars represent the 
identified dense cores.   The beam size is shown in the bottom 
left of the panel. 
Right: averaged spectrum of the H$^{13}$CO$^{+}$ line within 
the black circle showed on the left panel with different 
beam sizes. The two dashed vertical lines show the central 
velocities of the two velocity components measured at the native 
spatial resolution.  The beam sizes ($\theta$), velocity dispersion 
($\sigma_{\rm obs}$) and Mach number are presented inside the panel. 
\label{machmap}}
\end{figure*}

\subsection{Line width and Mach number}
\label{results:linewidth}

\subsubsection{$\rm H^{13}CO^{+}$}
\label{linewidthh13cop}
Single, double and triple velocity components 
were resolved in 84.8\%, 15.1\% and 1.1\% of the areas 
where significant H$^{13}$CO$^{+}$ emission was detected 
(Table \ref{tab:Mach}). 
We refer to spectra with single, double and triple velocity 
components as $v$1, $v$2 and $v$3, respectively. 
We found a $\langle \sigma_{\rm obs} \rangle$ = 0.31 km\,s$^{-1}$ 
mean  observed velocity dispersion\footnote{The one-dimensional 
observed velocity dispersion $\sigma_{\rm obs}$ is 
related to FWHM, define as 
$\sigma_{\rm obs} = \rm FWHM/(2\sqrt{2\, \rm ln\, 2})$.} 
from the dense cores, and a 
$\langle \sigma_{\rm obs} \rangle$ = 0.23 km\,s$^{-1}$ mean  
observed velocity dispersion exterior to the dense cores. 
The area of each dense core is marked with an open ellipse in 
Figure \ref{rgb}, while the remaining regions in the NGC 6334S 
are defined as areas exterior to the dense cores.  
The non-thermal velocity dispersion ($\sigma_{\rm nt}$)  of the 
H$^{13}$CO$^{+}$ line was estimated by 
\new{
$\sigma_{\rm nt} = \sqrt{\left(\sigma_{\rm obs}^{2} - \triangle_{\rm ch}^{2}/(2\sqrt{2\;\rm ln\;2})^{2} \right) - \sigma_{\rm th}^{2}}$,  }
where $\triangle_{\rm ch}$ and $\sigma_{\rm th}$ are 
velocity channel width and thermal velocity dispersion, 
respectively. 
The molecular thermal velocity dispersion was estimated 
by $\sigma_{\rm th}$ = $\sqrt{(k_{\rm B}T)/(\mu\, m_{\rm H})}$ = 
9.08 $\times$ 10$^{-2}$ km s$^{-1} 
\left(\frac{T}{\rm K}\right)^{0.5} \mu^{-0.5}$,
where $\mu = m/m_{\rm H}$ is the molecular weight, $m$ 
is the molecular mass, $m_{\rm H}$ is the proton mass, 
$k_{\rm B}$ is the Boltzmann constant, 
and $T$ is the gas temperature. 
The thermal velocity dispersion (sound speed $c_{\rm s}$) of 
the particle of mean mass was estimated assuming a mean mass 
of gas of 2.37$\, m_{\rm H}$ \citep{2008A&A...487..993K}.  
The gas temperatures, 
$T_{\rm NH_{3}}$, derived from the NH$_{3}$ line ratios 
were used to calculate the thermal velocity dispersion 
(Figure \ref{temperature}, see Section \ref{obs:fitting}). 
We assumed a gas kinetic temperature of 
$\langle T_{\rm NH_{3}} \rangle$ = 15~K  for regions where 
the NH$_{3}$ detection is insufficient for our temperature 
fittings or without NH$_{3}$ observations.  For each of the 
velocity components observed from the entire NGC 6334S region 
we also derived the Mach number, 
$\mathcal{M} = \sqrt{3} \sigma_{\rm nt}/c_{\rm s}$.
The derived Mach numbers range from 0.02 to 6.3, with a mean 
value of 1.5. 

Figure \ref{machmap} shows the Mach number map of NGC 6334S. 
From this map, it appears that non-thermal motions in the 
majority of the observed regions are subsonic or 
transonic ($0< \mathcal{M} < 2$).
The upper left panel of Figure \ref{machhist} shows a 
normalized histogram of Mach numbers, which is derived from the 
regions in which is detected one single velocity component.  We 
found that the $\sigma_{\rm nt}$ values of 29.6\%, 46.4\%, and 
24\% of these regions are subsonic ($\mathcal{M} \leqslant1$),
transonic ($1 < \mathcal{M} \leqslant2$), and supersonic 
($\mathcal{M} > 2$), respectively (for a complete summary 
see Table \ref{tab:Mach}). 

The upper right panel of Figure \ref{machhist} shows the 
histograms of Mach numbers derived from all observed velocity 
components.  The contribution from the areas of dense cores 
and from the areas exterior to the dense cores are 
distinguished with colors. 
From areas exterior to the dense 
cores, we observed a histogram similar to the one presented 
in the left panel of Figure \ref{machhist}. We found 
systematically higher yet mostly still subsonic-to-transonic 
Mach numbers in the areas of dense cores.  As can be seen 
from Figure \ref{machmap}, supersonic motions are mostly 
found in the central part of NGC 6334S, where there are multiple 
velocity components around dense cores.  In summary, the 
non-thermal motions in the NGC 6334S cloud and the embedded 
dense cores are predominantly subsonic and transonic.

\subsubsection{$\rm NH_{2}D$}
\label{linewidthnh2d}
\begin{figure*}[ht!]
\epsscale{1.3}
\plotone{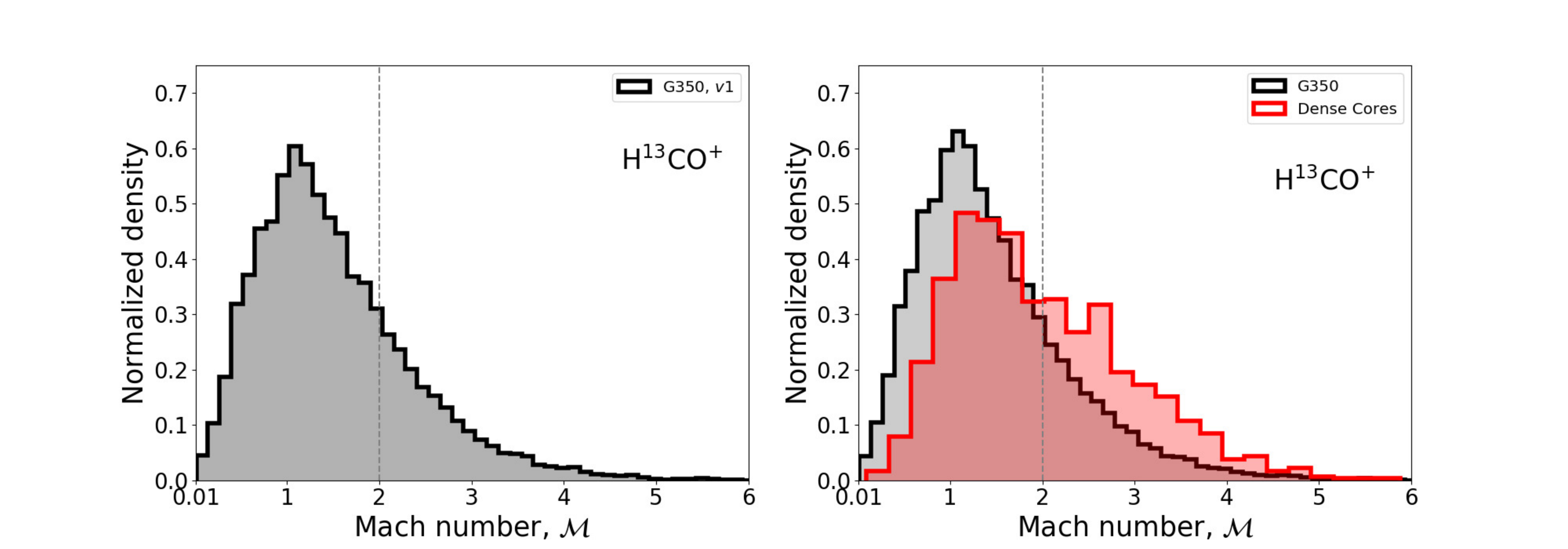}
\plotone{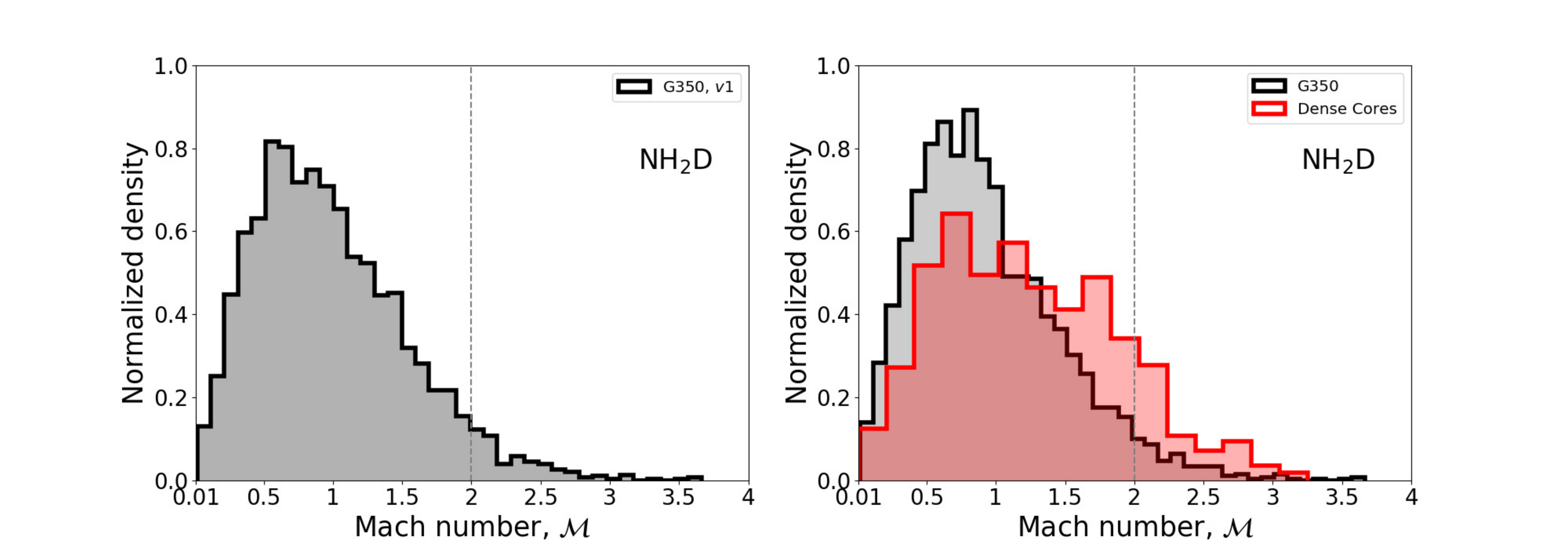}
\caption{ Gaussian kernel density estimate 
(KDE) of the Mach number distribution. 
Upper left: the  distribution of Mach numbers  for regions where 
a single velocity component was detected in NGC 6334S. 
Upper right: the distribution of Mach number for all velocity 
components for the entire NGC 6334S excluding the dense 
cores (black histogram), and for all velocity components from 
the dense cores only (red histogram).
Bottom: same as the upper panels but using the NH$_{2}$D line. 
The grey vertical dash-dotted line represents the position of Mach 
number equal to 2. 
\label{machhist}}
\end{figure*}

In 89.3\% and 10.7\% of the regions with NH$_{2}$D 
emission, we resolved single and double velocity components, 
respectively.  The mean $\sigma_{\rm obs}$ of the NH$_{2}$D line
is 0.21 km s$^{-1}$ and 0.16 km s$^{-1}$ for the dense cores and the 
remaining region in NGC 6334S, respectively, which are 
smaller than those of the H$^{13}$CO$^{+}$ line. 
Explanations for this discrepancy are discussed further in 
Section \ref{nh2ddecomp}. 
The Mach numbers estimated from the NH$_{2}$D line range from 
0.01 to 3.7, with a mean value of 1 (Table \ref{tab:Mach}).  The 
majority of cores (89\%) have a mean Mach number of less than 2, 
and 43\% of cores have a mean March number smaller than 1.  
Figure \ref{machhist} shows the normalized histogram of Mach 
numbers estimated from the the NH$_{2}$D line, which indicates 
that Mach numbers are smaller  than 2 at the majority of regions, 
and are between 2 and 3 in 4.9\% of the observed regions.

%

\begin{figure*}[ht!]
\epsscale{1.3}
\plotone{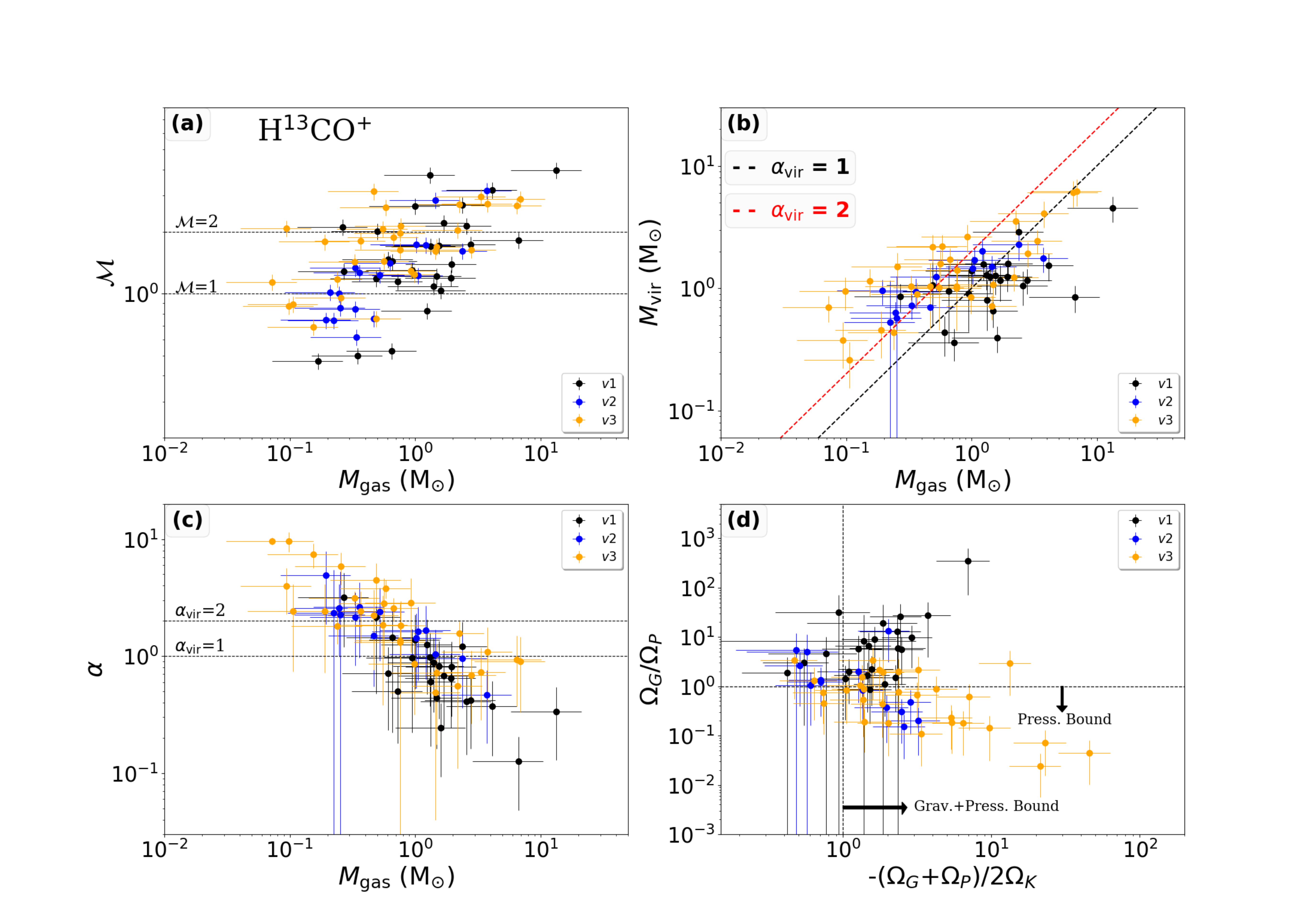}
\caption{Panel a: Mach number versus  gas mass for the 
H$^{13}$CO$^{+}$ decomposed structures. 
Panel b: virial mass versus  gas mass. The black and 
red dash-dotted lines show the $\alpha_{\rm vir}$ = 1 and 
$\alpha_{\rm vir}$ = 2 lines, respectively. 
Panel c: virial parameter versus  gas mass. 
Panel d: the ratio of $\Omega_{G}$/$\Omega_{P}$ 
versus ratio of -($\Omega_{G}$ +$\Omega_{P}$)/2$\Omega_{K}$.
$v1$, $v2$ and $v3$ represent the single, double and triple  
velocity components, respectively. 
\label{machvirh13cop}}
\end{figure*}

\begin{figure*}[ht!]
\epsscale{1.3}
\plotone{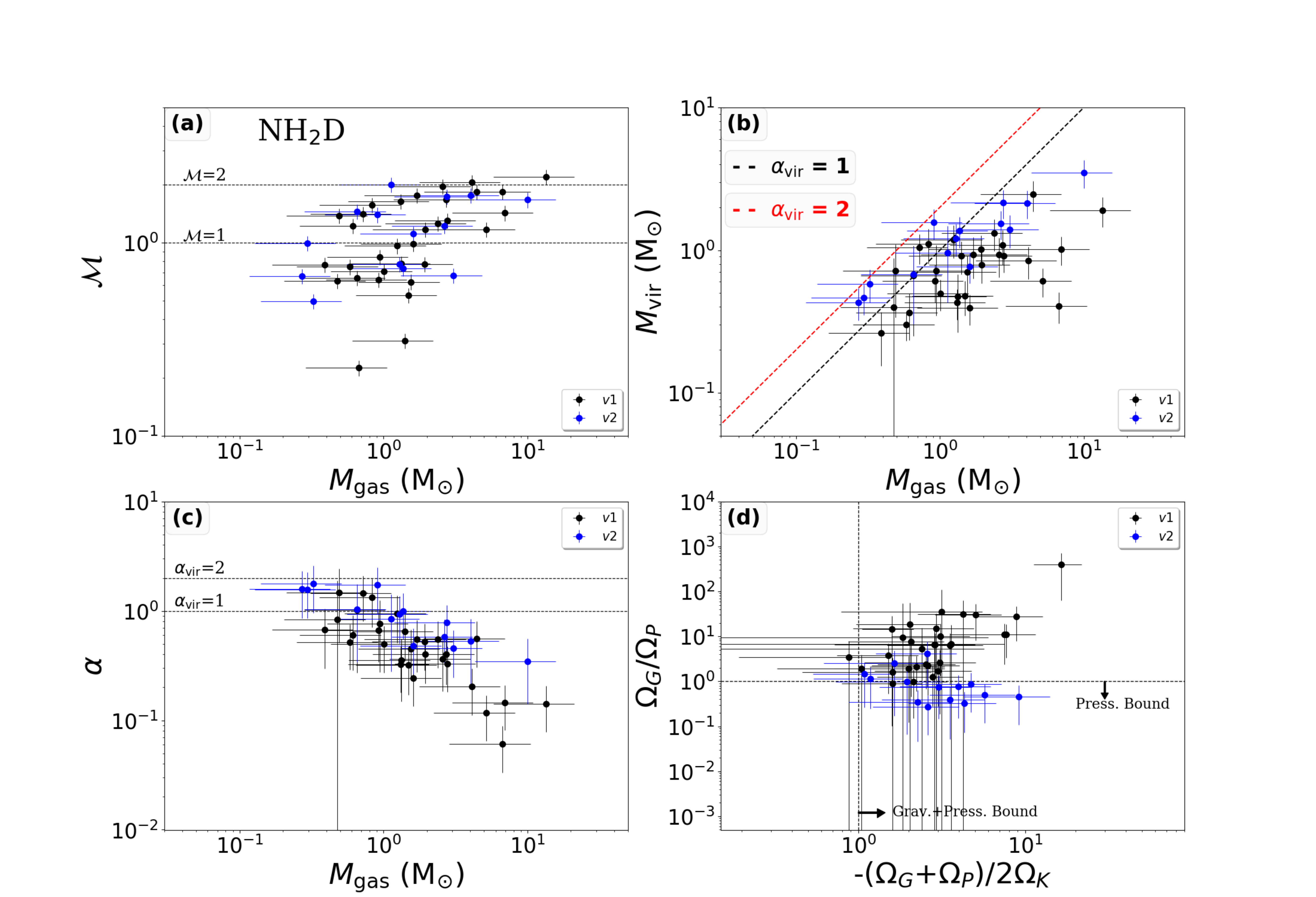}
\caption{ Same as Figure \ref{machvirh13cop} but for the NH$_{2}$D 
decomposed structures. 
\label{machvirnh2d}}
\end{figure*}


\subsection{Dense Core Properties}
\label{results:virial}
With the observed 3~mm continuum fluxes, we estimated 
the gas mass ($M_{\rm gas}$) of identified cores assuming 
optically thin modified black body emission in the Rayleigh-Jeans 
limit, following 
\begin{equation}
\label{dust_mass}
M_{\rm gas}=\eta \frac{F_{\nu} d^{2}}{B_{\nu}(T)\kappa_{\nu}},
\end{equation}
where $\eta$ = 100 is the assumed gas-to-dust mass ratio, 
$d$ is the source distance, 
$F_{\nu}$ is the continuum flux at frequency $\nu$, 
$B_{\nu}(T)$ is the Planck function at  temperature $T$, 
and $\kappa_{\nu}$ is the dust opacity at frequency $\nu$. 
We adopt $\kappa_{\rm 98.5\; GHz}$ = 0.235 cm$^{2}$g$^{-1}$, 
assuming  $\kappa_{\nu}$ = 10(${\nu}$/1.2 THz)$^{\beta}$ 
cm$^{2}$g$^{-1}$ and $\beta$ = 1.5 for all of dense cores  
\citep{1983QJRAS..24..267H}. 

The derived gas masses are between 0.17 and 14 \Mo: 
two cores have masses of 13 \Mo\ and 14 \Mo; the masses of 
12 cores are between 2 and 7 \Mo; the masses of the rest 
of the cores range from 0.17 to 1.97 \Mo\ (Table \ref{tab:cores}). 
Our 1$\sigma$ mass sensitivity corresponds to $\sim$0.03 \Mo.
The uncertainty in the continuum flux is adopted to be a typical 
value of 10\% in interferometer observations.  The 
uncertainty in distance from the trigonometric parallax 
measurement is about 20\% \citep{2014ApJ...784..114C}.  
The typical uncertainty in temperatures estimated from the 
NH$_{3}$ lines is about 15\%, which is mainly due to the low 
S/N in the NH$_{3}$ data.
$\eta$ is adopted to be 100 in this study, while its 
standard deviation is 23 (corresponding to 1$\sigma$ uncertainty 
of 23\%) assuming it is uniformly distributed between 70 and 
150  \citep{1990ApJ...359...42D,2003A&A...408..581V,
2017ApJ...841...97S}. We adopted a conservative uncertainty 
of 28\% in $\kappa_{\nu}$ 
\citep[e.g.,][]{2017ApJ...841...97S}.  Taking into account 
these uncertainties, we estimate an uncertainty of  57\%  
for the gas mass.

Assuming spherical symmetry for identified dense cores, we 
evaluated their volume densities 
($n_{\rm H_{2}}=M_{\rm gas}/(\mu m_{\rm H} 4 \pi R^{3}/3)$) 
using the gas mass and  effective 
radius.  The derived volume densities vary between 
4.3$\times$10$^{5}$ and 2.2$\times$10$^{8}$ cm$^{-3}$, with 
the mean and median value of  1.1$\times$10$^{7}$ and  
2.5 $\times$10$^{6}$ cm$^{-3}$, respectively. The typical 
uncertainty is about 58\% for volume density, with the 
exception of eight marginally resolved cores that have an 
uncertainty of $>$ 100\% due to a large uncertainty in 
the beam-deconvolved effective radius.

For cores which were resolved with multiple velocity 
components (Appendix \ref{appdecomposed}), 
we further estimated the gas mass ($M_{i}$) 
of each individual component $i$ assuming that its continuum flux 
is 
$F_{\rm cont}^{i} = 
F_{\rm cont}^{\rm tot} \times I_{\rm line}^{i}/I_{\rm line}^{\rm tot}$, 
where $F_{\rm cont}^{\rm tot}$ is the overall continuum flux, 
$I_{\rm line}^{i}$ is the velocity integrated intensity of the 
$i$th velocity component, and $I_{\rm line}^{\rm tot}$ = 
$\sum I_{\rm line}^{i}$ is the overall velocity integrated 
intensity. This provides a reasonable estimate of gas mass
for decomposed structures since the dense cores are dense 
enough ($> 4 \times 10^{5}$ cm$^{-3}$) for the gas and the 
dust to be well coupled \citep{2001ApJ...557..736G}. 
Using H$^{13}$CO$^{+}$ emission, we have decomposed 
 67 structures (hereafter decomposed structures). 

The virial ratio is evaluated for each decomposed  
structure using \citep{1992ApJ...395..140B}:
\begin{equation}
\label{equ:virial}
\alpha_{\rm vir} = \frac{M_{\rm vir}}{M_{\rm gas}}
\end{equation}

\begin{equation}
\label{equ:Mvir}
M_{\rm vir}=\frac{5}{a \beta} \frac{\sigma_{\rm tot}^{2} R}{G} ,
\end{equation}
where $M_{\rm vir}$ is the virial mass,  
$\sigma_{\rm tot}$ = $\sqrt{\sigma_{\rm nt}^{2} + c_{s}^{2}}$ is 
the total velocity dispersion, $R$ is the effective radius, 
$G$ is the gravitational constant, parameter $a$ equals to 
$(1-b/3)/(1-2b/5)$ for a power-law density profile 
$\rho \propto r^{-b}$ \citep{1992ApJ...395..140B}, 
and \new{ $\beta = (\arcsin e)/e $ } is the geometry factor. 
Here, we assume a typical density profile of $b$ = 1.6 for all 
decomposed structures  \citep{2002ApJ...566..945B,
2012ApJ...754....5B,2014ApJ...785...42P,2019ApJ...886..130L}.
The eccentricity $e$ is determined by the axis ratio of the dense 
structure,  $e = \sqrt{1 - f_{\rm int}^{2}}$.
The axis ratio of the decomposed structure is assumed to 
be the value of its corresponding dense core. 
Considering the projection effect and the dense cores are likely 
prolate ellipsoids \citep{1991ApJ...376..561M}, the observed axis 
ratio, $f_{\rm obs}$, is larger than the intrinsic axis ratio, $f_{\rm int}$.
The $f_{\rm int}$ can be estimated from $f_{\rm obs}$ with 
\begin{equation}
\label{ }
f_{\rm int} =  \frac{2}{\pi} \; f_{\rm obs} \; \mathcal{F}_{1} \; 
(0.5, 0.5, -0.5, 1.5, 1, 1 - f_{\rm obs}^{2})
\end{equation}
\citep{1983AJ.....88.1626F,2013ApJ...768L...5L}, where 
$\mathcal{F}_{1}$ is the Appell hypergeometric function of the 
first kind.

The corresponding virial mass of each velocity component is  
derived based on the corresponding effective radius, 
$R_{i} = R \times \sqrt{ A_{i}/A_{\rm core} }$. 
Here,  $R_{i}$ is the effective radius for the $i$th velocity 
component, $R$ is the dense core beam-deconvolved  effective 
radius (Table \ref{tab:cores}), $A_{i}$ is the number of pixels for 
the $i$th velocity component within the dense core measured in 
the moment-zeroth map,  and the $A_{\rm core}$ is the number 
of pixels for the dense core (Figure \ref{decomposed}).

Taking into account the uncertainties of $\sigma_{v}$ and $R$, 
leads to an uncertainty of about 24\% for virial mass. 
With the uncertainties of virial mass and dense core mass,  
the typical uncertainty is about 46\% for the virial ratios,
except for three marginal resolved dense cores that have an 
uncertainty of $>$ 100\% due to the large uncertainty in the 
beam-deconvolved effective radius.

Based on the 67 H$^{13}$CO$^{+}$ decomposed structures, 
we found that the virial ratios range from 0.1 to 9.7, 
with the mean and median value of 1.8 and 1.4, respectively. 
Forty-three out of 67 H$^{13}$CO$^{+}$ decomposed structures 
have virial ratios smaller than 2, while 27 of them have ratios  
below 1. Non-magnetized cores with $\alpha_{\rm vir} <$ 2,  
$\alpha_{\rm vir} \sim$ 1 and $\alpha_{\rm vir} <$ 1 are considered 
to be  gravitationally bound, in hydrostatic equilibrium and 
gravitationally unstable, respectively 
\citep{1992ApJ...395..140B,2013ApJ...779..185K}. 
Those with $\alpha_{\rm vir} >$ 2  could be gravitationally unbound. 
Figure \ref{machvirh13cop} shows the virial ratio 
versus the gas mass of the H$^{13}$CO$^{+}$ decomposed 
structures. From this plot, it is apparent that there is an inverse 
relationship between the mass and the virial ratio. 
This indicates that the most massive structures tend to be more 
gravitationally unstable, while the less massive  structures are 
gravitationally unbound and may eventually 
be disperse, unless there are other mechanism(s) that help to 
confine them such as external pressure  and/or embedded 
protostar(s) with a significant mass.

We have decomposed 45 structures with the NH$_{2}$D 
emission. For these NH$_{2}$D decomposed structures, 
the derived virial ratios are between 0.1 and 12, with the 
mean and median values of 1.3 and 0.6, respectively. All of 
the NH$_{2}$D decomposed structures have virial ratios 
lower than 2, and 36 of them have ratios smaller than 1.  
This suggests that these decomposed structures are 
gravitationally bound. The analysis of both NH$_{2}$D  and 
H$^{13}$CO$^{+}$ give a  similar result: the virial ratios of 
the decomposed structures decrease with increasing gas masses.

\subsection{Pressure}
\label{press}
Previous studies of star-forming regions have suggested that 
the external pressure provided by the ambient molecular 
cloud materials  might help to confine dense structures in 
molecular clouds  \citep{2006ApJ...646.1009K,
2013ApJ...770...44L,2017ApJ...846..144K,2015ApJ...805..185S,
2015MNRAS.450.1094P}. 
In order to examine whether the external pressure can play 
a significant role in confining the dense structures, 
we estimate the external pressure from the ambient cloud 
for each decomposed structure 
\citep{1989ApJ...345..782M,2017ApJ...846..144K} using: 
\begin{equation}
\label{pressure}
P_{\rm cl} = \pi G \bar{\Sigma}\Sigma_{r}  \beta_{\rm cl},
\end{equation}
where $P_{\rm cl}$ is the gas pressure, $\bar{\Sigma}$ 
is the mean column density of the cloud, 
$\Sigma_{r}$ is the column density integrated from 
the cloud surface to depth $r$, and $ \beta_{\rm cl}$ is the 
geometry factor for the cloud.  We assumed that $\Sigma_{r}$ 
is equal to half of the observed column density at the footprint of 
dense structures, 
$\Sigma_{r} =  \Sigma_{\rm obs}$/2, which is 
reasonable for most of star-forming regions 
\citep[e.g.,][]{2006ApJ...646.1009K,2017ApJ...846..144K}. 
To estimate the mean column density of a  cloud, 
we consider only the central region where 
the mean column density of the cloud is about
$\bar{\Sigma}$ = 0.15 g~cm$^{-2}$ estimated from the SED fitting 
(see Appendix \ref{SED}). 

The derived  external pressure of the H$^{13}$CO$^{+}$ 
decomposed structures is between  
9.1$\times 10^{6}$ and  1.9 $\times 10^{8} \, \rm K\, cm^{-3}$, 
with a mean value of 
$\langle P_{\rm cl}/k_{\rm B} \rangle =  3.8\times 10^{7} \, \rm K\, cm^{-3}$. 
For the NH$_{2}$D decomposed structures, the derived external 
pressures range from  9.1$\times 10^{6}$ to  
1.9$\times 10^{8} \, \rm K\, cm^{-3}$, 
with a mean value of 
$\langle P_{\rm cl}/k_{\rm B} \rangle =   3.9\times 10^{7} \, \rm K\, cm^{-3}$.

To determine the energy balance, we evaluated the 
external pressure energy, gravitational potential energy 
and internal kinetic energy in the virial equation 
\citep{1989ApJ...345..782M,1992ApJ...395..140B}: 
\begin{equation}
\label{Epressure}
\Omega_{P} = -4\pi P_{\rm cl} R^{3}
\end{equation}
\begin{equation}
\label{Egravitation}
\Omega_{G} = - \frac{3}{5}a \beta \frac{ G M^{2}}{R}
\end{equation}
\begin{equation}
\label{Ekinetic}
\Omega_{K} = \frac{3}{2} M \sigma_{\rm tot}^{2}
\end{equation}
where $\Omega_{P}$, $\Omega_{G}$ and $\Omega_{K}$ are the 
external pressure, gravitational and kinetic terms, respectively.

For the H$^{13}$CO$^{+}$ decomposed structure, 
the derived $\Omega_{P}$ ranges from 4.8$\times$10$^{40}$  to 
4.1$\times$10$^{44}$ erg, the $\Omega_{G}$ is between  
3.7$\times$10$^{40}$ and  7.0$\times$10$^{44}$ erg, and the 
$\Omega_{K}$ is between 3.7$\times$10$^{40}$ and  
1.3$\times$10$^{44}$ erg.  
The NH$_{2}$D decomposed structures have similar 
values compared to that of the H$^{13}$CO$^{+}$ with the 
$\Omega_{P}$ ranging 
from 1.5$\times$10$^{41}$  to 4.1$\times$10$^{44}$ erg, the 
$\Omega_{G}$  between 5.9$\times$10$^{41}$ and  
7.9$\times$10$^{44}$ erg, and the $\Omega_{K}$  between 
5.2$\times$10$^{41}$ and  5.6$\times$10$^{43}$ erg.

\section{Discussion}
\label{discussion}
%

\subsection{sonic-to-transonic motion}
\label{turbulence}
The mean observed values for  $\sigma_{\rm obs}$ are 
about 0.23 km s$^{-1}$ (or 0.54 km s$^{-1}$ in FWHM)
and 0.17 km s$^{-1}$  (or 0.40 km s$^{-1}$ in FWHM) 
for H$^{13}$CO$^{+}$ and NH$_{2}$D,  
respectively, which indicates that the 
observed spectra are resolved in these ALMA observations 
with a spectral velocity channel width of 0.21~km~s$^{-1}$ 
(see also the Appendix \ref{simulation}). 
These line widths may be regarded as conservative upper 
limits, given the limited spectral 
resolution, the hyperfine line splittings, the blending of 
velocity components along the line-of-sight, and the outflows 
and shocks all of which can lead to overestimates of the line 
widths \citep{2019ApJ...878...29L}.

In spite of the aforementioned biases, from 
Figure~\ref{machhist} 
we can  see that the spatially unresolved 
non-thermal motions are predominantly subsonic and transonic 
($\sim$77\% regions).  The normalized distribution function of 
$\mathcal{M}$ derived from the H$^{13}$CO$^{+}$ observations 
peaks at 1.2.  We have run a test with temperatures ranging from 8 
to 25 K, corresponding to the range of the  dust temperatures 
in the region, and found that  56\% to up to 99\% of Mach 
numbers  are smaller than 2.  This confirms that the Mach 
numbers are still dominated  by subsonic and transonic 
motions within this temperature range.

More interestingly, such features persist down to the dense 
core scales ($\sim$0.015 pc) but with a higher fraction of 
larger Mach numbers as shown in  Figure \ref{machhist}. 
There are several reasons that may  
cause  the relatively broad line widths toward the dense 
cores. First, the line widths may be broadened by protostellar 
activities (e.g., outflow, infall, and/or rotation). Second, given that 
the majority of dense cores are associated with multiple 
velocity components, the line widths could be enhanced by the 
intersection of gas components at different velocities. 
Third, high optical depths could broaden the detected 
line widths \citep{2016A&A...591A.104H}.  Unfortunately,  we 
can not discriminate between these possibilities with the 
data at hand. 

The subsonic-to-transonic dominated nature in both dense 
cores and NGC\,6334S cloud suggests that these cores are  
still at very early evolutionary stages such that the 
protostellar feedback-induced turbulence is small as compared 
to the initial turbulence. Such low turbulence  is similar to what 
is seen in low-mass star formation regions, e.g.,  Musca,  
L1517, NGC1333 and Taurus \citep{2011A&A...533A..34H,
2013A&A...554A..55H,2016A&A...587A..97H,2017A&A...606A.123H}

Figure \ref{machmap} shows the spectral lines of the 
H$^{13}$CO$^{+}$ at different spatial resolutions.  It is 
clear that the line widths increase with increasing beam 
sizes. 
Therefore, the observed supersonic non-thermal velocity 
dispersions in massive star forming regions often reported in 
the literature  
might be significantly biased by the poor spatial resolution,  
since the  spatially unresolved motions 
within the telescope beam can broaden the observed line widths.

\subsection{Dynamical State of Dense Cores}\label{dynamical}

The relationship of decreasing $\alpha_{\rm vir}$ with $M_{\rm gas}$ 
shown in Figures \ref{machvirh13cop} and \ref{machvirnh2d} has also 
been reported in previous studies of parsec scale massive 
clumps \citep{2015MNRAS.452.4029U,2013ApJ...779..185K}, 
subparsec scale ($\leqslant0.1$ pc) massive dense cores 
 \citep{2019ApJ...886..130L},   and low-mass star 
formation regions \citep{2008ApJ...672..410L,2009ApJ...696..298F}.
\citet{2018A&A...619L...7T} pointed out that such a decreasing 
trend of $\alpha_{\rm vir}$ may be attributed to systematic measurement 
errors when a particular molecular line preferentially traces 
molecular gas with densities above the critical density of the 
line transition. This effect could lead to a  lower estimate of 
the virial parameter due to an underestimation of the observed 
line width. However, this effect may not be  significant 
since the critical density of a given gas tracer is typically higher 
than the effective excitation density because of the effect of 
radiative trapping 
\citep{2015PASP..127..299S,2019FrASS...6....3H}. 
The assumption of a certain density profile (equation \ref{equ:Mvir}, 
with $b = 1.6$) could introduce some uncertainties to the virial 
mass, while it does not appear to be the dominant factor in the 
inverse $\alpha_{\rm vir} - M_{\rm gas}$ relation since the density 
structures do not show significant differences between 
the massive cores and the low-mass cores   
\citep[see for example,][]{2000ApJ...537..283V,2001A&A...365..440M,
2002ApJS..143..469M,2002ApJ...575..337S,2002ApJ...566..945B,
2012ApJ...754....5B,2014ApJ...785...42P,2015ApJ...798..128T,
2019ApJ...886..130L}.

Chemical segregation may introduce biases in the 
measurements of $M_{\rm vir}$. For example, the  
study of the B-type star-forming region G192.16-3.84 
\citep{2013ApJ...771...71L} found that the H$^{13}$CO$^{+}$  
emission did not trace the gravitationally accelerated, 
high-velocity gas that is probed by the  H$^{13}$CN emission 
and other hot core tracers.  The decreasing trend 
of $\alpha_{\rm vir}$ with $M_{\rm gas}$ therefore may be 
alternatively interpreted as indicating   
that higher mass cores evolve faster, and have lower 
H$^{13}$CO$^{+}$ and NH$_{2}$D abundances at their 
centers.  This hypothesis can be tested by spatially resolving 
gas temperature profiles for individual cores or  
 by spatially resolving the abundance distribution of 
H$^{13}$CO$^{+}$ and NH$_{2}$D at higher angular resolutions. 
Additionally, \cite{2019MNRAS.490.3061V} suggest that the core 
sample satisfies the relation $M_{\rm gas} \propto R^{p}$, 
with $0 < p < 3$, would naturally yield an inverse 
$\alpha_{\rm vir} - M_{\rm gas}$ relation.

On the other hand, investigations of a protocluster-forming 
molecular clump IRDC G28.34+0.06 have revealed that the line 
width decreases toward smaller spatial scales of denser regions 
\citep{2008ApJ...672L..33W}. The change of line width 
could be due to dissipation of turbulence from the 
clump scale to core scale \citep{2008ApJ...672L..33W,
2015ApJ...804..141Z}. The spatially resolved studies 
of the OB cluster-forming molecular clump 
G33.92+0.11 \citep{2012ApJ...756...10L,2015ApJ...804...37L,
2019ApJ...871..185L} have demonstrated that in its inner 
$\sim$0.5 pc region, the turbulent energy is indeed much less 
important as compared to the gravitational potential energy. 
Molecular gas in this region is highly gravitationally 
unstable, is collapsing towards the central ultra compact 
\HII\ region, and is fragmenting to form a cluster of young 
stellar objects (YSOs).  The 0.5 pc scale 
molecular clump and all embedded objects are detected with 
small $M_{\rm vir}$ and have $\alpha_{\rm vir}<$ 1, which is 
likely due to the fact that the dominant rotational and infall motions 
are perpendicular to the line-of-sight. 
Compared with our present observations on NGC 6334S, we 
hypothesize that turbulence in this case can dissipate more 
rapidly in cores with higher masses and higher densities. 
The diagnostics of the dynamical state of dense 
cores may require taking the projection effect into 
consideration; this can be tested by spatially resolving the 
gas motions within the identified cores.

\subsubsection{{\rm H$^{13}$CO$^{+}$} decomposed structures}

Two observational biases can lead to an overestimate of the 
virial ratio: (1) overestimating the line width due to poor 
fitting; (2) incorrectly accounting for potential multiple 
velocity components not resolved in 
the current observations, which would broaden the observed 
line width. The observed velocity dispersions are similar 
among the identified cores with a mean value of 0.31 km s$^{-1}$, 
with the exception of one massive and one intermediate-mass 
cores having slightly larger observed velocity dispersion of 
around 0.5 km s$^{-1}$. To address the first issue, we 
inspected the line fitting of the H$^{13}$CO$^{+}$ line for 
each core and removed or refitted the pixels with poor fitting 
in an effort to insure that the high virial ratio we measure 
in the less massive regimes is real.  The 
second bias does not appear to be a significant factor in our dataset 
since the spectral resolution is high enough to resolve the 
observed lines (see the Appendix \ref{simulation}). We conclude 
that the dense structures with high virial ratios are indeed unbound 
gravitationally.

The structures with high virial ratios may disperse and may  
not form stars in the future  without additional 
mechanism(s) to counteract the internal pressures. 
In Section \ref{results:virial}, our simplified virial 
analysis considered only balance between self-gravity and 
the internal support, and did not include the external 
pressure. In order to check whether the external pressure 
term plays an important role in confining the identified 
dense structures, we have computed the energy due to the 
external pressure ($\Omega_{P}$), gravitational potential 
energy ($\Omega_{G}$) and internal kinetic energy 
($\Omega_{K}$) (see Section \ref{press}).

When $\Omega_{P} > \Omega_{G}$ the external pressure 
plays a dominant role in binding the dense structure, while 
$\Omega_{P} < \Omega_{G}$ means that the gravity is the  
dominant element.  In the panel c of Figure 
\ref{machvirh13cop}, we show the ratio of 
$\Omega_{G}$/$\Omega_{P}$ versus the ratio of 
-($\Omega_{G}$ +$\Omega_{P}$)/2$\Omega_{K}$. 
The mean ratio of $\Omega_{G} / \Omega_{P}$ is 8.9, with 
a median value of 1.4. From Figure \ref{machvirh13cop}, 
it is clear that the gravity dominates over the external pressure 
($\Omega_{G} / \Omega_{P} >$  1) for more than half 
(60\%) of the structures, while 20 structures are dominated by 
the external pressure  ($\Omega_{G} / \Omega_{P} <$  1).

In the energy density form, the virial ratio is  
-2$\Omega_{K}$/($\Omega_{G}$ +$\Omega_{P}$), when the external 
pressure is considered.  In Figure \ref{machvirh13cop}, the 
vertical dashed line marks the locus of virial equilibrium. 
The structures that lie to the right of this line are confined
by gravity and external pressure, while structures on 
the left of this line are unbound. The mean energy ratio of 
-($\Omega_{G}$ +$\Omega_{P}$)/2$\Omega_{K}$ is about 3.6, 
with a median value of  1.8.  As described in the previous 
paragraph (Section \ref{results:virial}), 
a fraction (36\%) of decomposed structures are gravitationally 
unbound in the simplified virial analysis, while most of them 
(79\%) can be confined by gravity and external pressure, 
-($\Omega_{G}$ +$\Omega_{P}$)/2$\Omega_{K} >$ 1. This suggests 
that the external pressure plays a role in confining the dense 
structures in this early evolutionary stage of star formation 
region. There are 14 structures which are not bound, even when 
accounting for the effect of external pressure.

In the above analysis, the external pressure only includes 
the pressure from the clump-scale cloud, but all of our dense 
cores are embedded within filamentary structures in the 
NGC 6334S cloud. These filaments could provide an additional 
external pressure on these dense cores. We are not able to 
accurately compute the filament pressure because it is 
difficult to extract the column density belonging to the 
filaments from this complex system. \cite{2017ApJ...846..144K} 
suggests that the filament pressure is about a 
factor of 10 smaller than the pressure from the larger-scale 
cloud; however, this is a lower limit for their objects 
because they used parameters for low-mass star formation 
regions to estimate the filament pressure.  
The filament pressure in our sources could be 
higher than 10\% of the pressure from the larger-scale cloud. 
In addition, the infall ram pressure is expected to provide 
an additional support to confine the dense cores 
\citep{2009ApJ...704.1735H,2017ApJ...846..144K}.

\subsubsection{{\rm NH$_{2}$D} decomposed structures}
\label{nh2ddecomp}
We also analyzed the energy density ratio for the NH$_{2}$D 
decomposed structures as shown in  the panel d of 
Figure \ref{machvirnh2d}. These ratios behave in a similar way as  
in the H$^{13}$CO$^{+}$ decomposed structures. The gravity 
plays a bigger role than the external pressure in confining 31 out 
of 45 structures with the other 14 structures being bound 
mostly due to the external pressure. 
In general, the virial ratio ($\alpha_{\rm vir}$ and 
-2$\Omega_{K}$/($\Omega_{G}$ +$\Omega_{P}$)) of the NH$_{2}$D 
decomposed structures tends to be lower than those of the 
H$^{13}$CO$^{+}$, due to the fact that the widths of the NH$_{2}$D 
line are narrower than those of the H$^{13}$CO$^{+}$ line. 
The NH$_{2}$D emission traces colder gas and is less affected by the 
protostellar feedback (e.g., molecular outflows) as compared 
with the H$^{13}$CO$^{+}$ emission.  
The derived line widths are also affected by the hyperfine 
splitting due to the limited spectral resolution. 
Overall, the results from both NH$_{2}$D and H$^{13}$CO$^{+}$ 
lines are consistent, which supports our conclusion that the 
dense cores and their parent cloud are dominated by subsonic 
and transonic motions with external pressure helping to 
confine the younger and less massive dense structures.

\section{Conclusion}
\label{conclusion}
We conducted ALMA and VLA observations toward the massive 
IRDC NGC 6334S. We used the cold and dense gas 
tracers, H$^{13}$CO$^{+}$, NH$_{2}$D and NH$_{3}$,  to 
study the gas properties from the clump scale down to the 
dense core scale. The main findings are
\begin{enumerate}
  \item  Forty-nine embedded dense cores are identified in  
  the ALMA 3~mm dust continuum image. The majority of the 
  cores are located in the central region of NGC 6334S and are 
  embedded in filamentary structures. The masses of the dense 
  cores range from 0.17  to 14 \Mo, with the effective 
  radius between 0.005 and 0.041 pc. 
  
  \item The non-thermal velocity dispersion reveals that this 
  massive cloud, as well as the embedded dense cores, are 
  dominated by subsonic and transonic  motions. 
  The narrow non-thermal line widths resemble those seen in 
  the low-mass star formation regions. 
  Supersonic non-thermal velocity dispersions have been 
  reported often in massive star forming regions. However, we 
  caution that these studies might be significantly biased by 
  poor spatial resolutions that broaden the observed line 
  widths due to unresolved motions within the telescope beam.

  \item The majority of identified dense cores have  
  multi-velocity components. We find that the virial 
  ratios decrease with increasing mass despite the fact that 
  they are dominated by subsonic and transonic motions. This 
  implies that most massive structures tend to be gravitationally 
  unstable. Most of dense structures can be confined by the gravity 
  and external pressure, although a fraction of them are unbound. 
  This result indicates that the external pressure plays a 
  role in confining the dense structures during the early stages 
  of star formation. 
\end{enumerate}


\acknowledgments
We thank an anonymous referee for constructive comments that help 
improving this paper.
This work is supported by the National Key R\&D Program of China 
(No. 2017YFA0402604) and the Natural Science Foundation of China 
unders grant 11590783 and 11629302. 
S.L. acknowledges support from the CfA pre-doctoral fellowship 
and Chinese Scholarship Council. 
H.B. acknowledges support from the European Research Council 
under the Horizon 2020 Framework Program via the ERC 
Consolidator Grant CSF-648505. 
A.P. acknowledges financial support from CONACyT and 
UNAM-PAPIIT IN113119 grant, M\'exico. 
J.M.G.  acknowledges support from MICINN AYA2017-84390-C3-2-R. 
ALMA is a partnership of ESO (representing its member states), NSF (USA) 
and NINS (Japan), together with NRC (Canada) and NSC and ASIAA (Taiwan), 
in cooperation with the Republic of Chile. 
The Joint ALMA Observatory is operated by ESO, AUI/NRAO and NAOJ.

\vspace{5mm}
\facilities{ALMA, JVLA, APEX, \texttt{Planck},
\texttt{Herschel}.}

\software{ 
APLpy \citep{2012ascl.soft08017R}, 
Astropy \citep{2013A&A...558A..33A}, 
CASA \citep{2007ASPC..376..127M},
Matplotlib \citep{Hunter:2007},
Numpy \citep{2011CSE....13b..22V}, 
PySpecKit \citep{2011ascl.soft09001G},
SAOImageDS9 \citep{2003ASPC..295..489J}.
}

\bibliographystyle{aasjournal}
\bibliography{ms}


\begin{longrotatetable}
\begin{deluxetable*}{ccccccccccccc}
\tablecaption{Dense core properties
\label{tab:cores}}
\tablehead{
\colhead{Core ID} & \colhead{R.A.} &  \colhead{Decl.} &  \colhead{$\sigma_{maj}$} &  \colhead{$\sigma_{min}$} 
&  \colhead{PA} &  \colhead{$R_{\rm eff}$} &  \colhead{$R_{\rm deceff}$} &  \colhead{$S_{\nu}^{\rm peak}$} &  \colhead{$F_{\nu}$} 
&  \colhead{$M_{\rm gas}$} &  \colhead{$T$} &  \colhead{$n_{\rm H_{2}}$} \\  
\colhead{} & \colhead{(hh:mm:ss)} &   \colhead{(dd:mm:ss)} &   \colhead{(arcsec)} &   \colhead{(arcsec)} &   \colhead{(deg)} 
&   \colhead{(arcsec)} & \colhead{(arcsec)} &   \colhead{(mJy beam$^{-1}$)} &   \colhead{(mJy)} &   \colhead{($M_{\odot}$)} 
&   \colhead{(K)} &   \colhead{(cm$^{-3}$)}  
} 
\startdata
1 & 17:19:07.41 & -36:07:14.16 & 13.51  & 9.39  & 124.46  & 6.76  & 6.49  & 1.213 & 1.846  & 14.03  & 18  & 9.27E+05 \\
2 & 17:19:08.85 & -36:07:03.59 & 7.24  & 4.53  & 97.12  & 3.44  & 2.93  & 3.770 & 1.491  & 13.36  & 15  & 9.72E+06 \\
3 & 17:19:05.92 & -36:07:21.43 & 6.55  & 3.85  & 77.05  & 3.02  & 2.45  & 2.771 & 0.837  & 6.93  & 16  & 8.69E+06 \\
4 & 17:19:08.34 & -36:06:56.14 & 9.41  & 5.85  & 54.56  & 4.46  & 4.06  & 0.896 & 0.595  & 4.47  & 18  & 1.23E+06 \\
5 & 17:19:09.39 & -36:06:31.18 & 5.17  & 3.19  & 79.69  & 2.44  & 1.69  & 2.896 & 0.574  & 5.19  & 15  & 1.97E+07 \\
6 & 17:19:06.68 & -36:07:17.62 & 6.95  & 6.51  & 72.60  & 4.04  & 3.61  & 1.032 & 0.560  & 5.46  & 14  & 2.14E+06 \\
7 & 17:19:05.23 & -36:07:50.55 & 3.98  & 2.62  & 81.06  & 1.94  & 0.83  & 3.749 & 0.467  & 6.67  & 10  & 2.15E+08 \\
8 & 17:19:08.20 & -36:07:10.00 & 5.05  & 3.13  & 68.55  & 2.39  & 1.58  & 2.197 & 0.417  & 4.08  & 14  & 1.92E+07 \\
9 & 17:19:08.36 & -36:04:31.70 & 5.16  & 4.15  & 67.48  & 2.78  & 2.12  & 1.134 & 0.307  & 2.77  & 15  & 5.37E+06 \\
10 & 17:19:05.54 & -36:06:33.64 & 10.60  & 7.49  & 88.96  & 5.35  & 5.05  & 0.320 & 0.304  & 3.40  & 13  & 4.84E+05 \\
11 & 17:19:06.03 & -36:07:16.98 & 4.96  & 3.38  & 80.35  & 2.46  & 1.72  & 1.343 & 0.270  & 1.80  & 20  & 6.48E+06 \\
12 & 17:19:09.23 & -36:09:10.58 & 6.89  & 5.21  & 95.40  & 3.60  & 3.12  & 0.614 & 0.263  & 2.38  & 15  & 1.44E+06 \\
13 & 17:19:12.92 & -36:06:13.95 & 8.99  & 6.94  & 160.36  & 4.74  & 4.31  & 0.334 & 0.252  & 2.28  & 15  & 5.22E+05 \\
14 & 17:19:04.08 & -36:07:18.75 & 5.24  & 4.17  & 38.86  & 2.81  & 2.05  & 0.878 & 0.231  & 1.96  & 16  & 4.15E+06 \\
15 & 17:18:59.99 & -36:07:19.56 & 4.43  & 3.48  & 78.65  & 2.36  & 1.54  & 0.927 & 0.229  & 2.61  & 12  & 1.32E+07 \\
16 & 17:18:59.63 & -36:07:27.08 & 5.02  & 3.22  & 80.09  & 2.42  & 1.66  & 0.711 & 0.220  & 1.72  & 17  & 6.90E+06 \\
17 & 17:19:09.27 & -36:06:24.85 & 6.76  & 3.63  & 91.57  & 2.98  & 2.37  & 0.737 & 0.218  & 1.97  & 15  & 2.73E+06 \\
18 & 17:19:06.20 & -36:10:18.96 & 7.08  & 5.25  & 95.22  & 3.66  & 3.20  & 0.464 & 0.213  & 1.93  & 15  & 1.08E+06 \\
19 & 17:19:06.04 & -36:07:11.91 & 9.76  & 5.58  & 88.28  & 4.43  & 4.06  & 0.287 & 0.188  & 1.48  & 17  & 4.07E+05 \\
20 & 17:19:04.93 & -36:06:51.23 & 3.93  & 3.45  & 87.31  & 2.21  & 1.17  & 1.087 & 0.178  & 1.61  & 15  & 1.83E+07 \\
21 & 17:19:05.93 & -36:10:24.93 & 7.68  & 3.79  & 84.00  & 3.24  & 2.69  & 0.459 & 0.172  & 1.56  & 15  & 1.47E+06 \\
22 & 17:19:04.88 & -36:07:16.13 & 4.82  & 3.05  & 86.72  & 2.30  & 1.48  & 0.605 & 0.165  & 1.33  & 16  & 7.56E+06 \\
23 & 17:19:11.91 & -36:07:01.46 & 4.82  & 4.09  & 25.94  & 2.67  & 1.80  & 0.929 & 0.165  & 1.49  & 15  & 4.70E+06 \\
24 & 17:19:09.52 & -36:07:00.40 & 5.65  & 5.61  & 88.28  & 3.38  & 2.82  & 0.426 & 0.163  & 1.25  & 17  & 1.02E+06 \\
25 & 17:19:06.43 & -36:10:29.03 & 6.21  & 6.08  & 178.28  & 3.69  & 3.18  & 0.315 & 0.156  & 1.41  & 15  & 8.02E+05 \\
26 & 17:19:08.25 & -36:07:01.92 & 4.97  & 3.57  & 16.54  & 2.53  & 1.09  & 0.689 & 0.147  & 1.31  & 15  & 1.87E+07 \\
27 & 17:19:09.13 & -36:06:36.59 & 5.84  & 3.50  & 88.28  & 2.71  & 2.05  & 0.467 & 0.115  & 1.86  & 9  & 3.95E+06 \\
28 & 17:19:03.91 & -36:10:06.01 & 6.59  & 4.57  & 61.61  & 3.30  & 2.76  & 0.225 & 0.114  & 1.03  & 15  & 9.01E+05 \\
29 & 17:19:12.42 & -36:08:58.99 & 8.43  & 3.35  & 122.53  & 3.19  & 2.02  & 0.234 & 0.111  & 1.00  & 15  & 2.23E+06 \\
30 & 17:19:06.44 & -36:10:40.93 & 5.92  & 4.58  & 18.95  & 3.13  & 2.40  & 0.254 & 0.104  & 0.94  & 15  & 1.25E+06 \\
31 & 17:19:04.90 & -36:07:55.00 & 5.47  & 4.48  & 103.28  & 2.98  & 2.35  & 0.311 & 0.092  & 2.54  & 15  & 1.17E+06 \\
32 & 17:19:05.76 & -36:06:26.55 & 4.94  & 4.30  & 91.33  & 2.77  & 2.09  & 0.317 & 0.081  & 0.72  & 15  & 1.46E+06 \\
33 & 17:19:05.87 & -36:05:38.76 & 3.99  & 3.11  & 95.33  & 2.12  & 1.03  & 0.417 & 0.080  & 0.72  & 15  & 1.22E+07 \\
34 & 17:19:04.90 & -36:06:47.57 & 4.08  & 2.31  & 88.28  & 1.84  & ... & 0.651 & 0.074  & 0.67  & 15  & ... \\
35 & 17:19:07.70 & -36:09:03.24 & 3.54  & 2.38  & 87.15  & 1.74  & ... & 0.259 & 0.072  & 0.65  & 15  & ... \\
36 & 17:19:12.78 & -36:06:03.37 & 5.25  & 4.42  & 119.33  & 2.89  & 2.20  & 0.719 & 0.072  & 0.66  & 15  & 1.14E+06 \\
37 & 17:19:05.13 & -36:07:44.19 & 6.42  & 3.82  & 79.90  & 2.97  & 2.39  & 0.236 & 0.069  & 0.89  & 11  & 1.19E+06 \\
38 & 17:19:00.22 & -36:07:06.41 & 4.09  & 3.53  & 6.75  & 2.28  & ... & 0.320 & 0.066  & 0.84  & 11  & ... \\
39 & 17:19:05.41 & -36:07:18.64 & 3.95  & 2.90  & 86.25  & 2.03  & 0.97  & 0.467 & 0.064  & 0.58  & 15  & 1.18E+07 \\
40 & 17:19:09.35 & -36:06:51.91 & 4.62  & 3.43  & 101.73  & 2.39  & 1.53  & 0.331 & 0.063  & 0.49  & 17  & 2.51E+06 \\
41 & 17:19:00.65 & -36:07:17.19 & 4.97  & 3.68  & 149.83  & 2.57  & 1.26  & 0.243 & 0.062  & 0.59  & 14  & 5.40E+06 \\
42 & 17:19:06.47 & -36:06:25.96 & 5.71  & 2.85  & 88.60  & 2.42  & 1.57  & 0.284 & 0.056  & 0.53  & 14  & 2.52E+06 \\
43 & 17:19:06.36 & -36:07:36.21 & 4.13  & 3.00  & 46.38  & 2.11  & ... & 0.373 & 0.055  & 0.50  & 15  & ... \\
44 & 17:19:05.44 & -36:07:08.49 & 4.86  & 2.92  & 96.22  & 2.26  & 1.32  & 0.285 & 0.049  & 0.27  & 23  & 2.18E+06 \\
45 & 17:19:03.99 & -36:07:01.38 & 4.37  & 2.59  & 79.76  & 2.02  & 0.97  & 0.313 & 0.043  & 0.39  & 15  & 7.74E+06 \\
46 & 17:19:03.28 & -36:06:06.66 & 3.80  & 2.98  & 104.60  & 2.02  & ... & 0.256 & 0.039  & 0.35  & 15  & ... \\
47 & 17:19:04.29 & -36:06:48.94 & 3.49  & 2.99  & 166.63  & 1.94  & ... & 0.233 & 0.029  & 0.27  & 15  & ... \\
48 & 17:19:02.21 & -36:07:16.60 & 2.78  & 2.25  & 91.49  & 1.50  & ... & 0.266 & 0.020  & 0.18  & 15  & ... \\
49 & 17:19:06.34 & -36:08:19.64 & 2.78  & 2.38  & 90.09  & 1.55  & ... & 0.233 & 0.019  & 0.17  & 15  & ... \\
\enddata
\tablecomments{
 R.A.: right ascension. 
 Decl.: declination. $\sigma_{maj}$: beam-convolved major axis. 
$\sigma_{min}$: beam-convolved minor axis. PA: position angle. 
$R_{\rm eff}$: beam-convolved effective radius. 
$R_{\rm deceff}$: beam-deconvolved effective radius.  
$F_{\nu}^{\rm peak}$: peak intensity. 
$F_{\nu}$: total integrated flux. $M_{\rm gas}$: gas mass. 
$T$: temperature. $n_{\rm H_{2}}$: averaged volume density. 
}
\end{deluxetable*}
\end{longrotatetable}

%
\begin{deluxetable*}{cccccCrlc}[b!]
\tablecaption{Mach number 
\label{tab:Mach}}
\tablecolumns{6}
\tablewidth{0pt}
\tablehead{
\colhead{Line} & \colhead{Velocity component} 
& \colhead{Area} 
& \colhead{$\mathcal{M}$} 
& \colhead{$\mathcal{M}_{\rm mean}$} 
& \colhead{$\mathcal{M}_{\rm median}$} 
& \colhead{$\mathcal{M} \leqslant $ 1} 
& \colhead{1 $ < \mathcal{M} \leqslant $ 2} & \colhead{$\mathcal{M} > $ 2}
}
\startdata
\multirow{4}{*}{H$^{13}$CO$^{+}$} 
 & $v1$ & 0.72  & 0.02-6.3 & 1.5 & 1.4 & 29.6\% & 46.4\% & 24\%  \\
 & $v2$ & 0.13  & 0.04-5.4 & 1.5 & 1.2 & 32.9\% & 47.5\% & 19.6\%  \\
 & $v3$ & 0.01   & 0.1-5.1 & 1.4 &  1.2 & 32.5\% & 48.6\% & 18.9\% \\
 & All  & 0.86   & 0.02-6.3 & 1.5 &  1.3 & 30.1\% & 46.6\% & 23.3\% \\
\hline  
\multirow{3}{*}{NH$_{2}$D} 
 & $v1$ & 0.25 & 0.01-3.7  & 1.0  & 0.9 & 57.8\% & 37.5\% & 4.6\%  \\
 & $v2$ & 0.03  & 0.05-3.0  & 1.0 & 0.8 & 61.2\% & 31.6\% & 7.2\%    \\
 & All & 0.28  & 0.01-3.7  & 1.0 & 0.9 & 58.2\% & 36.9\% & 4.9\%    \\
\enddata
\tablecomments{ 
Area: the projection area in pc$^{2}$.
$v_{\rm 1}$: single velocity component. 
$v_{\rm 2}$: double velocity components.
$v_{\rm 3}$: triple velocity components. 
All: all emission components.
}
\end{deluxetable*}




\clearpage 

\appendix

\section{Column density and dust temperature}
\label{SED}
We adopted the image combination and SED fitting procedure 
presented in \cite{2016ApJ...828...32L,2017ApJ...840...22L}. 
The procedure is briefly summarized here.
\texttt{Planck} 353 GHz emission data are deconvolved using 
Lucy-Richardson algorithm with a model map from 
\texttt{Herschel}  \texttt{PACS} 160 \um\ (level2.5), 
\texttt{SPIRE} 250/350/500 \um\ (level3) derived 870 \um\ 
data. Then the deconvolved data were linearly combined 
with the ATLASGAL 870 \um\ data in the Fourier domain.
\texttt{PACS} 160 \um, \texttt{SPIRE} 250 \um\ and the 
combined 870 \um\ data are used to fit the column density 
$N(\rm H_{2})$ and dust temperature $T_{d}$,  
achieving a resolution of 18\arcsec.

We performed an SED fit of each pixel, using a 
modified-black-body model:
\begin{equation}
\label{sedeq}
I_{\nu} = B_{\nu}(T_{d})(1-e^{-\tau_{\nu}})
\end{equation}
where $\tau_{\nu}$ = $\mu m_{\rm H} N(\rm H_{2})  \kappa_{\nu}$, 
$m_{\rm H}$ is the mean molecular weight and we adopt a value 
of 2.8.    
We adopted a dust opacity law of 
$\kappa_{\nu}$ = 10(${\nu}$/1.2 THz)$^{\beta}$ cm$^{2}$g$^{-1}$, 
where $\kappa_{\rm 353 GHz}$ = 1.6 is the dust opacity per unit 
mass (gas and dust) at reference frequency 353 GHz and $\beta$ 
is fixed to 1.5 \citep{1983QJRAS..24..267H}.

\section{Decomposed structures}
\label{appdecomposed}
\begin{figure*}[ht!]
\epsscale{1.1}
\plotone{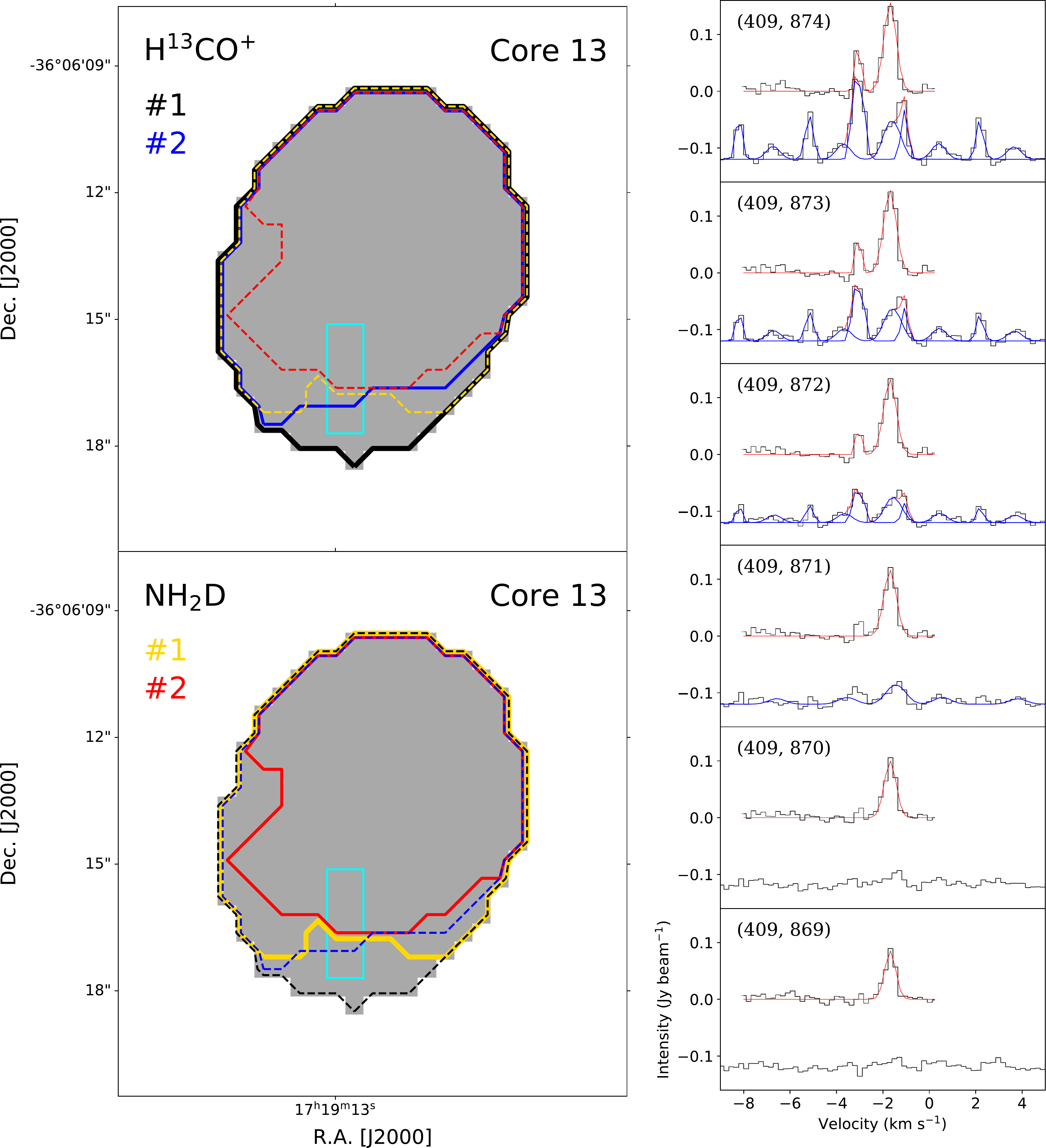}
\caption{Left: schematic view of decomposed structures identified 
using the H$^{13}$CO$^{+}$ (upper left) and NH$_{2}$D (bottom left) 
lines for dense core 13. The greyscale is the dense core revealed 
by the continuum emission. The black and blue contours represent 
the spatial distributions of the H$^{13}$CO$^{+}$ decomposed structures, 
while the yellow and red contours represent the spatial distributions 
of the NH$_{2}$D decomposed structures. 
The contours are derived by the emission regions of corresponding 
velocity components, while \#1 and \#2 represent the first and 
second velocity components, respectively. 
Right: spectra of the H$^{13}$CO$^{+}$ (upper) and NH$_{2}$D 
(lower) lines within the cyan rectangle shown on the left panel. 
For the H$^{13}$CO$^{+}$ spectrum, the solid red curve represents 
the fitted result. 
For the NH$_{2}$D spectrum, the solid blue curve represents the 
components that are fitted, while solid red curve is the sum 
of them. The complete figure set (15 images) is available in 
the online journal.
\label{decomposed}}
\end{figure*}

As mentioned in Section \ref{results:virial}, a dense core identified 
in dust continuum emission can 
be decomposed into sub-structures using the H$^{13}$CO$^{+}$ 
and NH$_{2}$D lines. 
Figure \ref{decomposed} shows the spatial distributions of the 
decomposed structures and their spectral line profile. 
The estimations of gas mass and effective radius of the decomposed 
structures are described in Section \ref{results:virial}.

\begin{figure*}[ht!]
\epsscale{1.}
\plotone{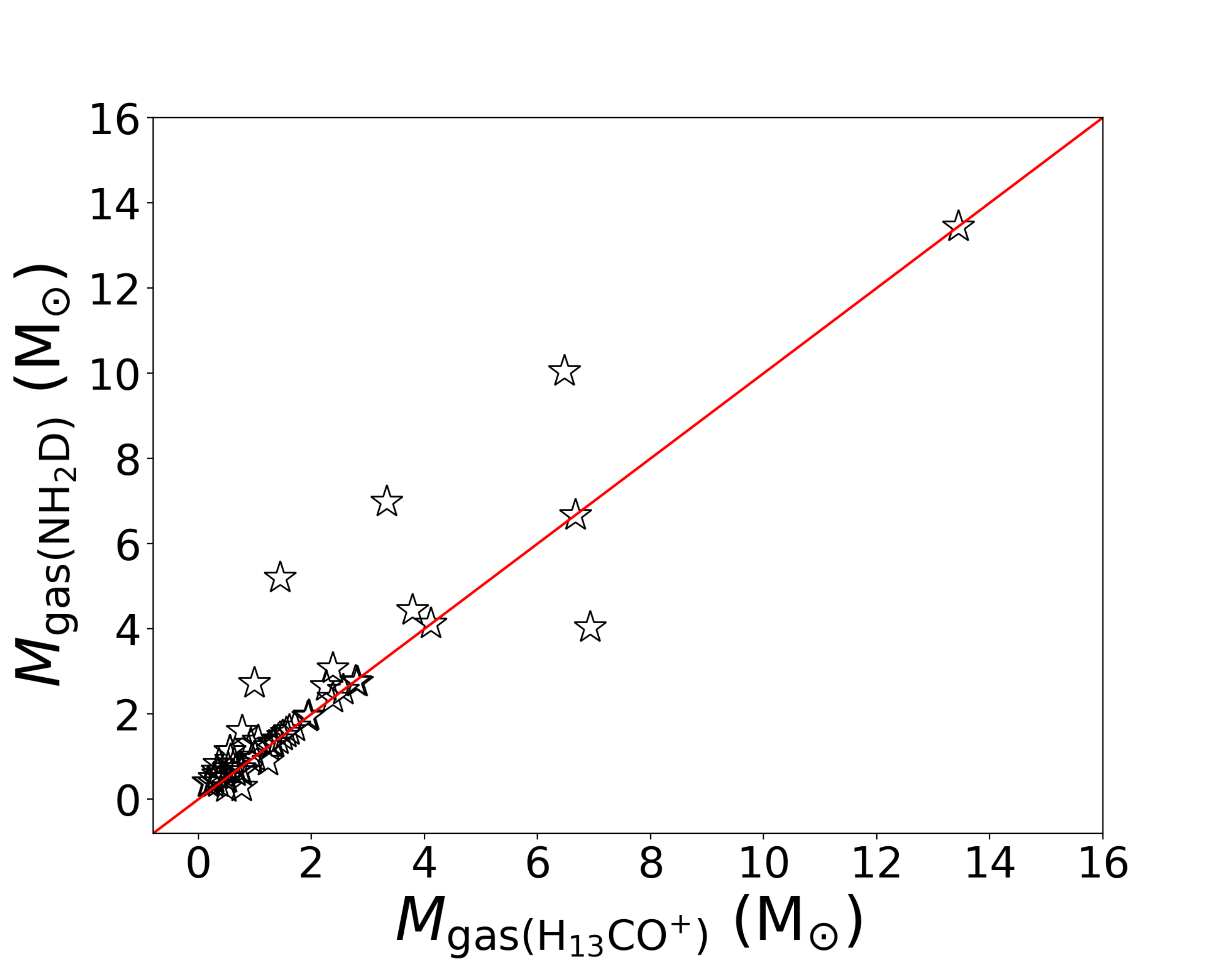}
\caption{The gas masses of the H$^{13}$CO$^{+}$ decomposed 
structures versus that of the NH$_{2}$D decomposed structures. 
The solid red line is the y = x line. 
\label{destructure}}
\end{figure*}

The majority of decomposed structures identified using the 
H$^{13}$CO$^{+}$ and NH$_{2}$D lines have similar gas masses 
and have significant overlap in projection (Figures \ref{decomposed} 
and \ref{destructure}), 
which indicates that they are consistent with each other.   
Some of the H$^{13}$CO$^{+}$ and NH$_{2}$D decomposed 
structures partly overlap in projection; this is due to the discrepancy 
in the emission regions  of the H$^{13}$CO$^{+}$ and NH$_{2}$D lines, 
especially where the lines present multiple velocity components. 
This discrepancy is probably due to the presence of faint velocity 
components that are limited by the sensitivity of our observations. 
On the other hand, the H$^{13}$CO$^{+}$ and NH$_{2}$D may  
trace slightly different gas components toward some dense cores, 
since they have different excitation conditions and chemical properties.

The detected multiple velocity components in a line profile could be 
attributed to optical depth effects, outflows, or a number of dense 
cores along the line of sight \citep{2013ApJ...771...24S}. 
Optical depth effects should not be the dominant mechanism 
causing the multiple velocity components in the line profile, since 
both H$^{13}$CO$^{+}$ and NH$_{2}$D emission are generally optically thin, 
although these lines might become optically thick in the densest 
regions of the dense core. 
The narrow line widths seen in both H$^{13}$CO$^{+}$ and NH$_{2}$D 
lines suggest that they are tracing quiescent rather than the shocked material. 
Therefore, the detected multiple velocity components in both 
H$^{13}$CO$^{+}$ and NH$_{2}$D lines are most likely due to a number 
of dense cores along the line of sight \citep[e.g.,][]{2013ApJ...771...24S}.

\section{Simulation}
\label{simulation}
\begin{figure*}[ht!]
\epsscale{1.35}
\plotone{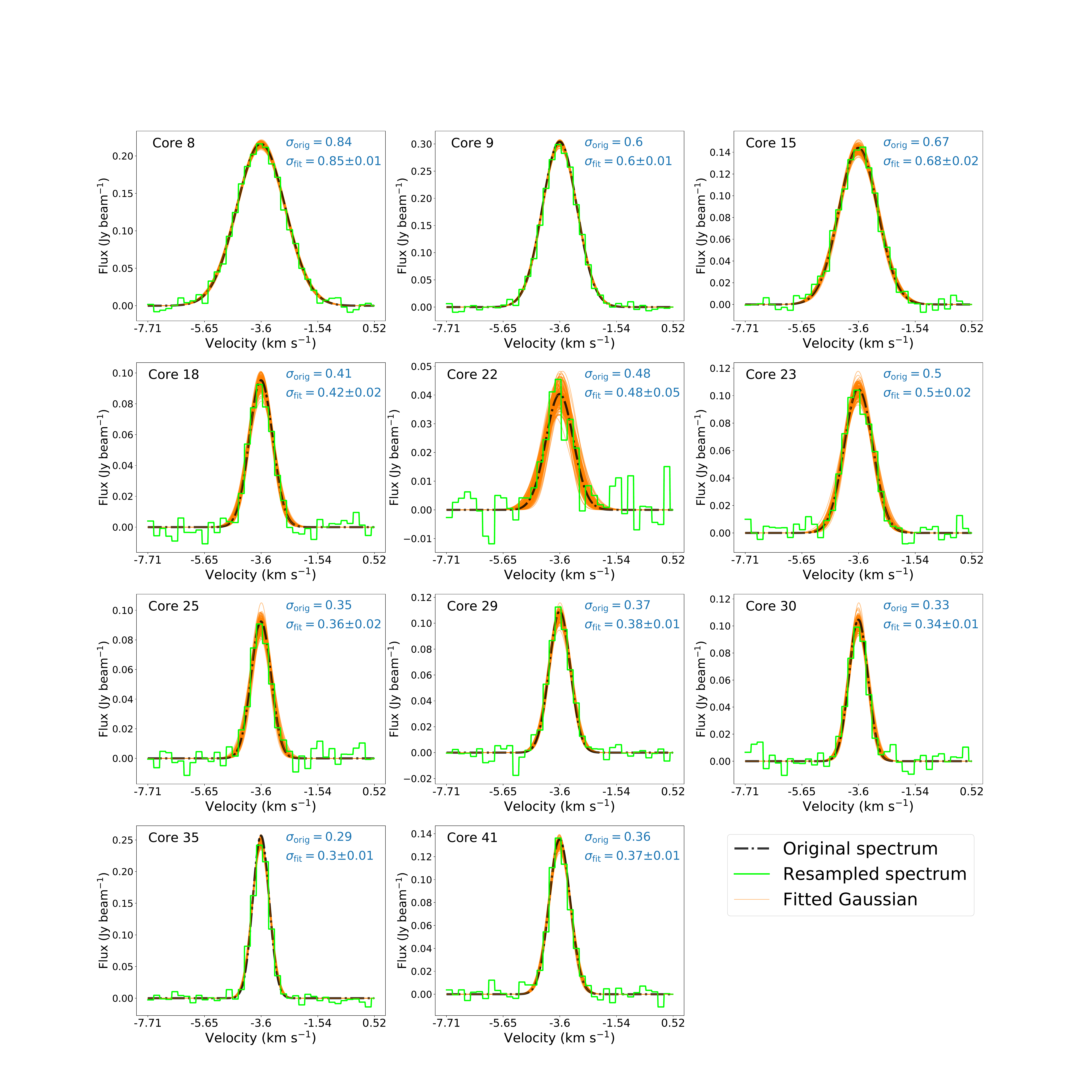}
\caption{The black line is the original synthesized spectrum in the 
spectral resolution of 0.013 km s$^{-1}$, while the green line is the 
original synthesized spectrum smoothed to 
a spectral resolution of 0.21 km s$^{-1}$. The yellow lines are fitted 
results for each iteration. The dense core ID is presented 
on the upper left panel.  
The synthesized spectral line width and  fitted line width are presented 
on the upper right panel. 
\label{simul_spec}}
\end{figure*}

To examine whether the observed line widths are  biased by 
the spectral resolution or not, we constructed a simplified model. 
Assuming that the dense cores are in virial equilibrium 
($\alpha_{\rm vir}$ = 1), 
the virial velocity dispersion ($\sigma_{\rm vir}$) can be described by  
\begin{equation}
\label{virdisp}
\sigma_{\rm vir} = \sqrt{\frac{\alpha_{\rm vir}\; M_{\rm gas}\; G}{5 R}}
\end{equation}
where $M_{\rm gas}$ and $R$ are the gas mass and effective 
radius. We used the observed line intensity and $\sigma_{\rm vir}$ 
as input parameters to synthesize a spectrum with a spectral 
resolution of 0.013 km s$^{-1}$ for  dense cores 
(black lines in Figure \ref{simul_spec}). 
We fit Gaussian line profiles to the synthetic spectra, that 
have been smoothed to a spectral resolution of 0.21 km s$^{-1}$ 
and with injected a random noise in 1000 iterations; we only show 
the dense cores that exhibit a single velocity component in the 
H$^{13}$CO$^{+}$ emission. 
Figure \ref{simul_spec} shows the fitting results. 
The fitted line width agrees with the input line width, indicating 
that the measured line widths 
are not significantly affected by the current spectral resolution in 
this study.  We ran the same test for the NH$_{2}$D line, and 
reached  the same conclusion. 

\clearpage
\newpage

\begin{center}
Presented here are individual plots for Figure~\ref{decomposed} 
which are available as the online material. The black cross marks 
the position of the spectra. 
\end{center}
\begin{tabular}{cc}
\includegraphics[clip,width=1\hsize]{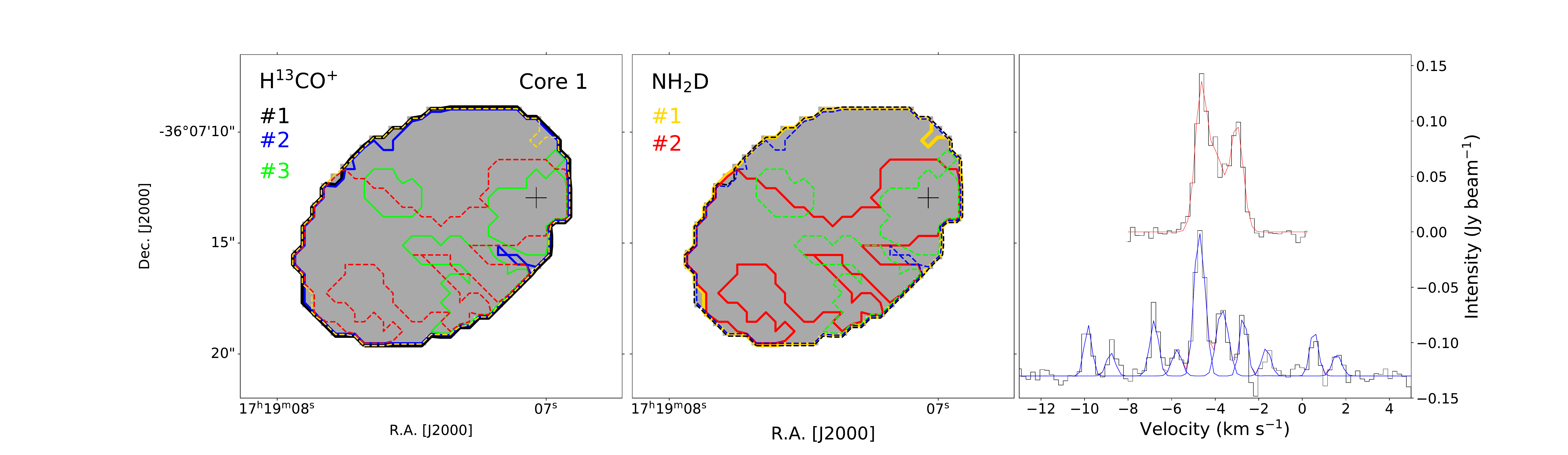}  \\
\includegraphics[clip,width=1\hsize]{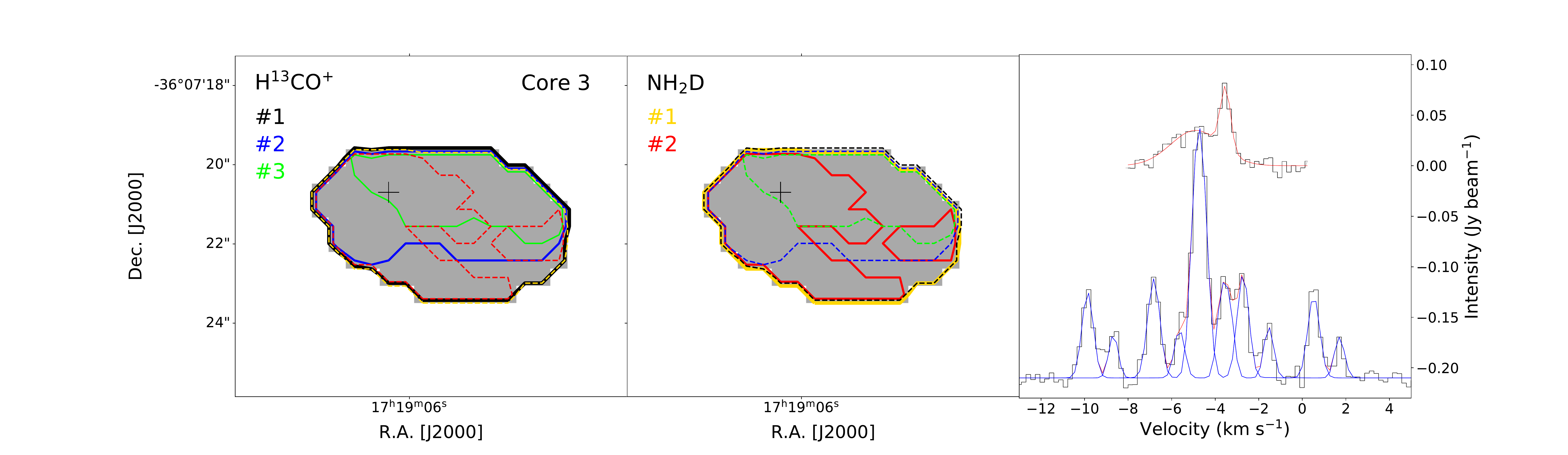}  \\
\includegraphics[clip,width=0.75\hsize]{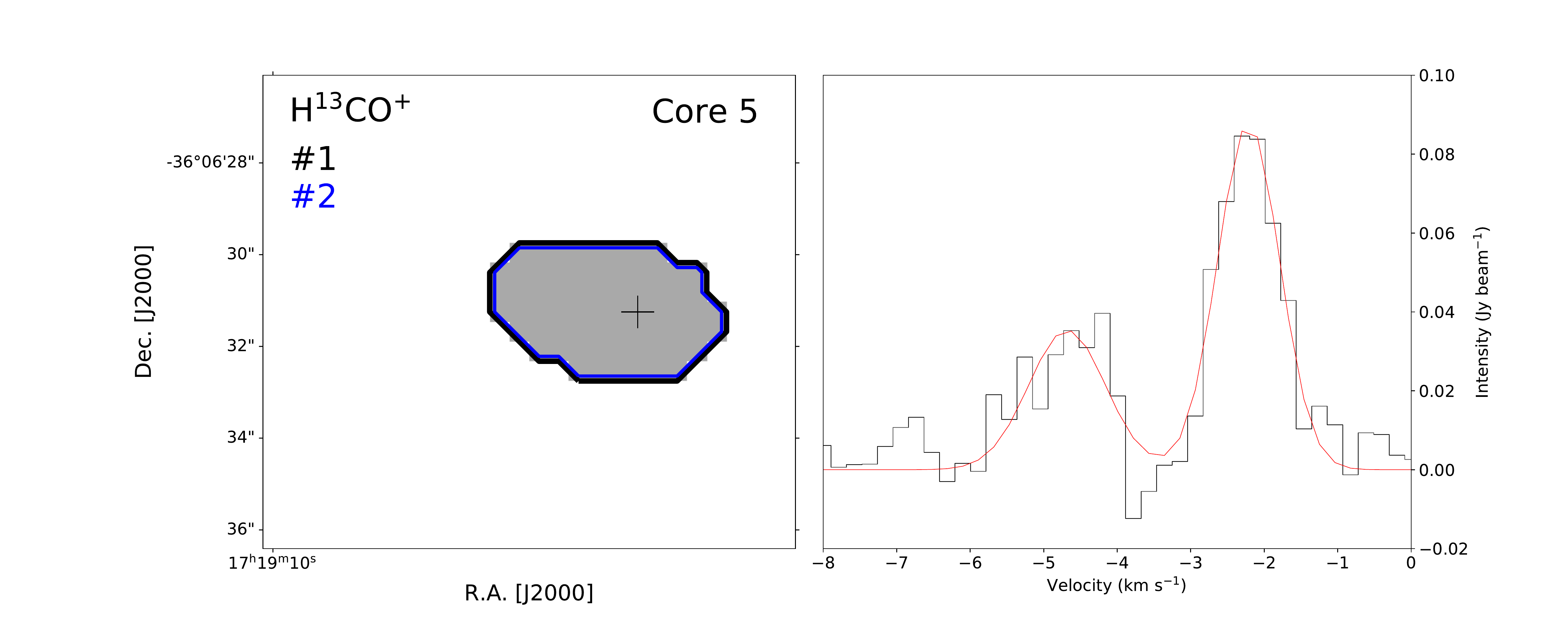}  \\
\includegraphics[clip,width=1\hsize]{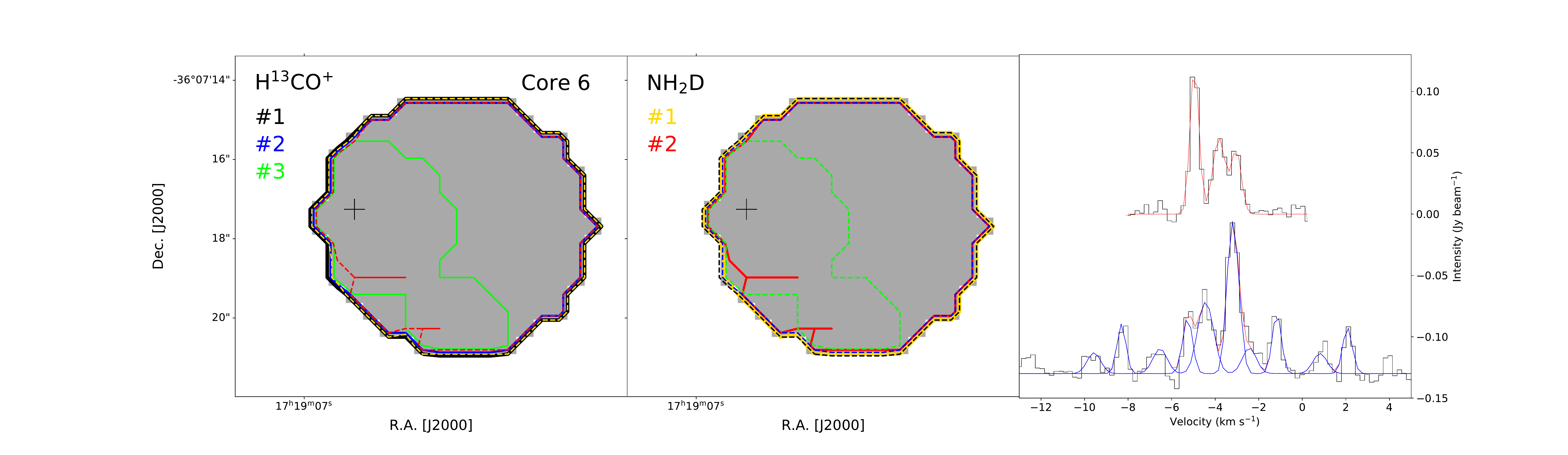} 
\end{tabular}

\newpage

\begin{center}
Presented here are individual plots for Figure~\ref{decomposed} 
which are available as the online material. 
\end{center}
\begin{tabular}{cc}
\includegraphics[clip,width=1\hsize]{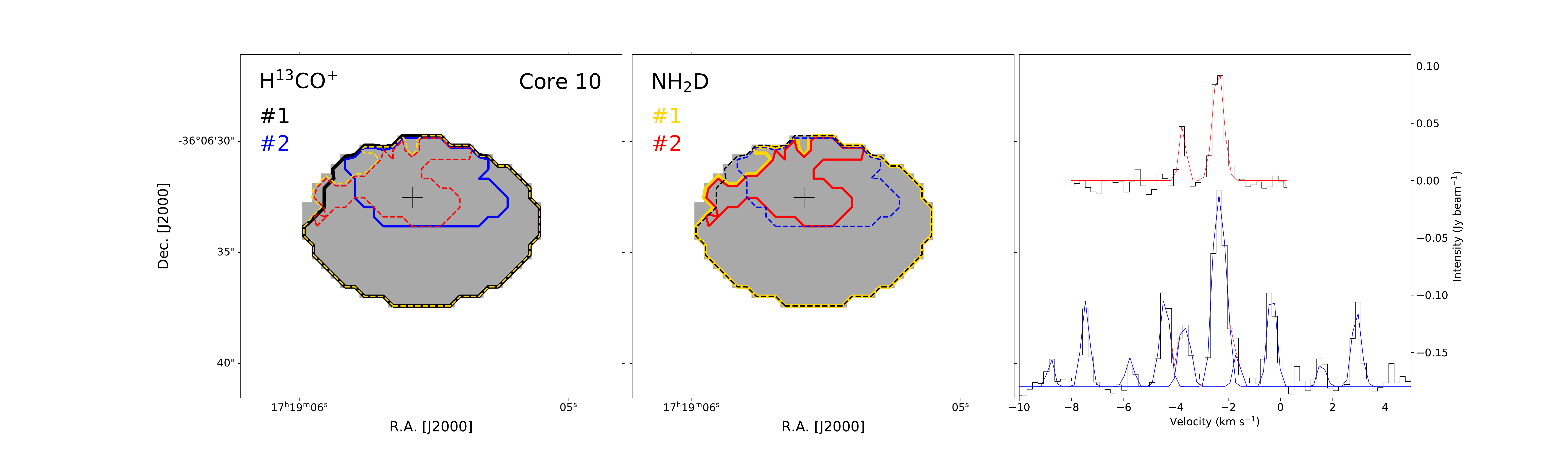}  \\
\includegraphics[clip,width=1\hsize]{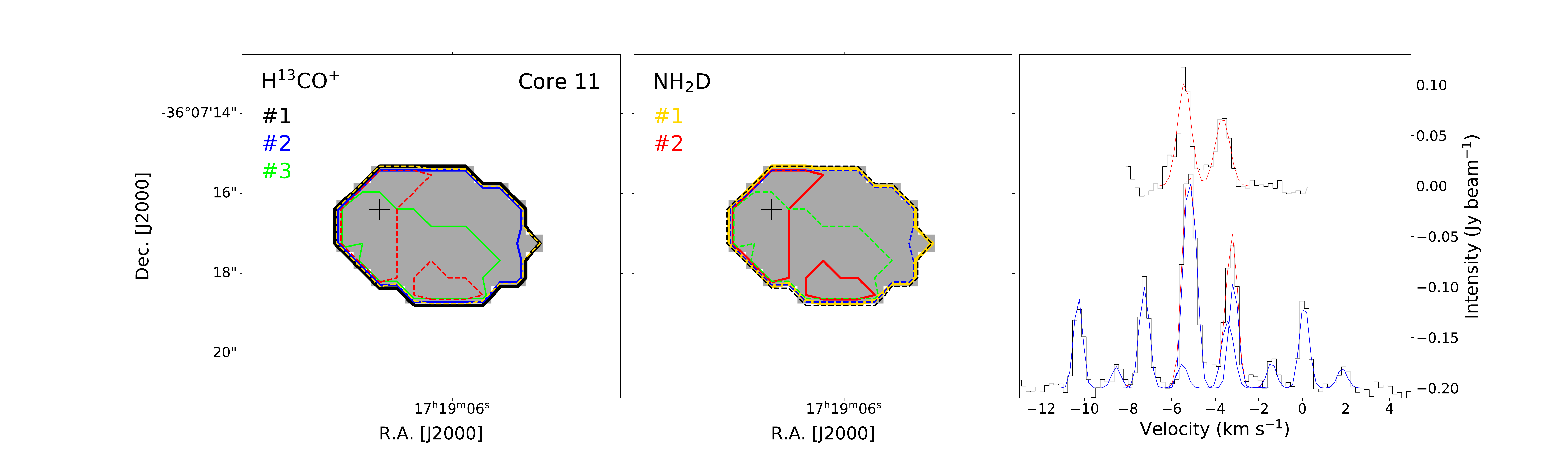}  \\
\includegraphics[clip,width=0.7\hsize]{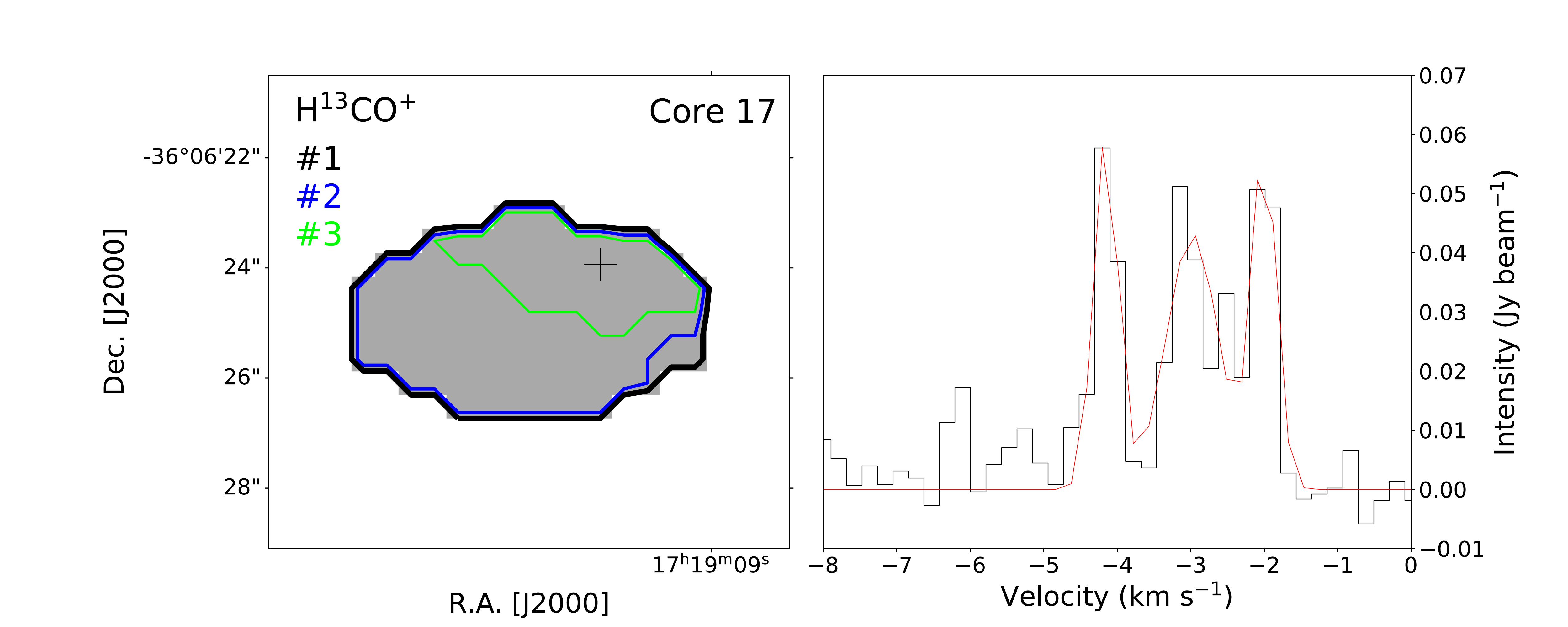}  \\
\includegraphics[clip,width=1\hsize]{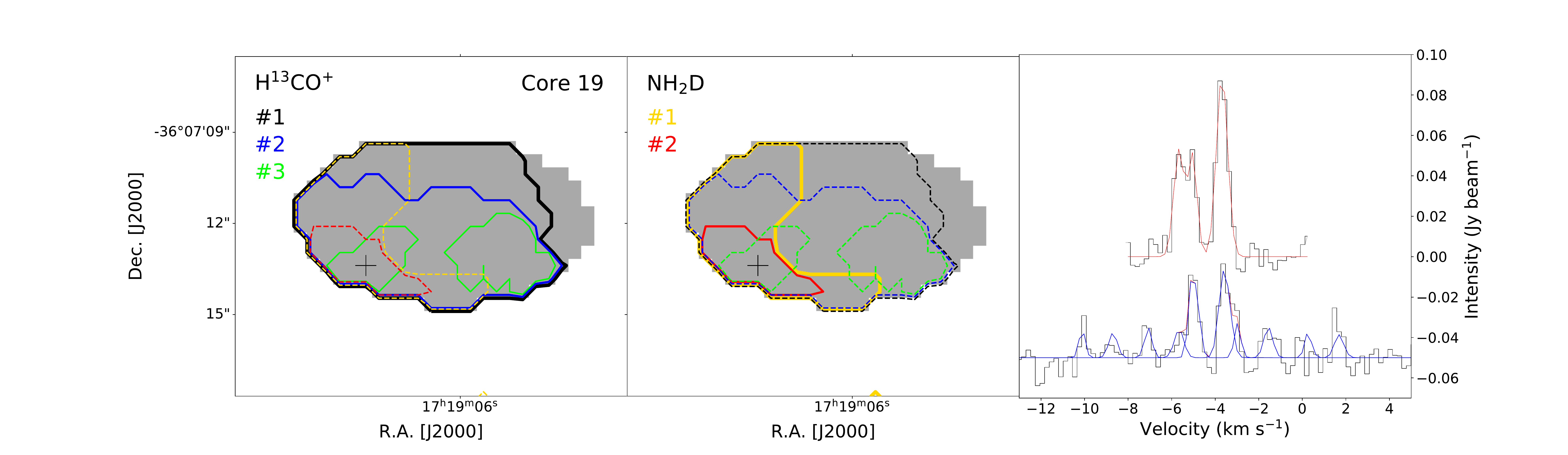} 
\end{tabular}

\newpage

\begin{center}
Presented here are individual plots for Figure~\ref{decomposed} 
which are available as the online material. 
\end{center}
\begin{tabular}{c}
\includegraphics[clip,width=1\hsize]{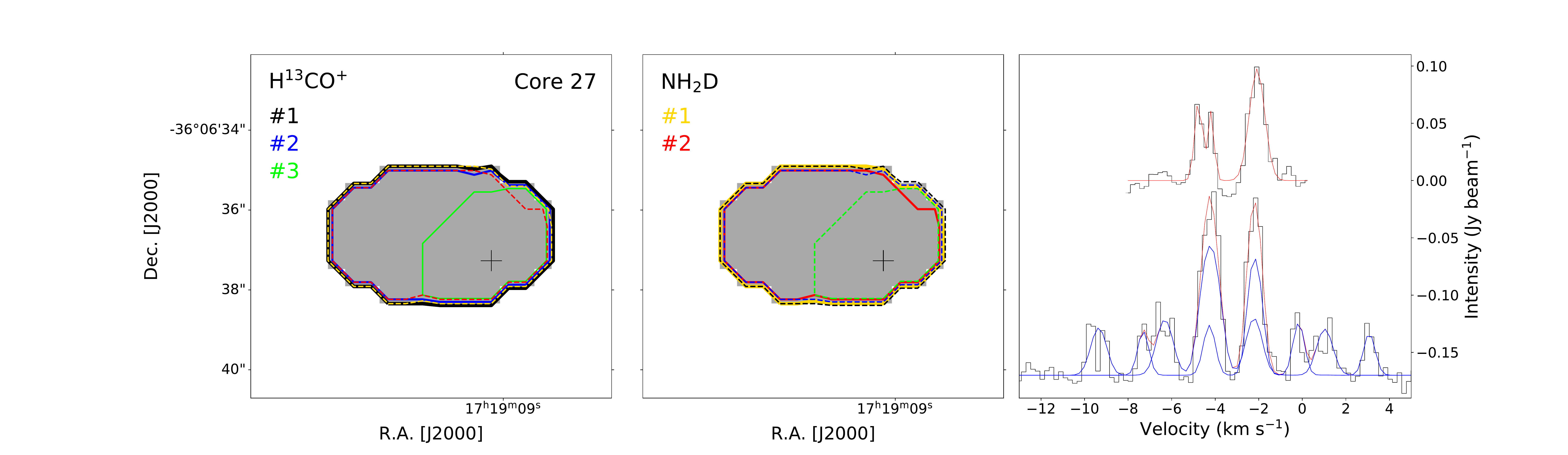}  \\
\includegraphics[clip,width=0.76\hsize]{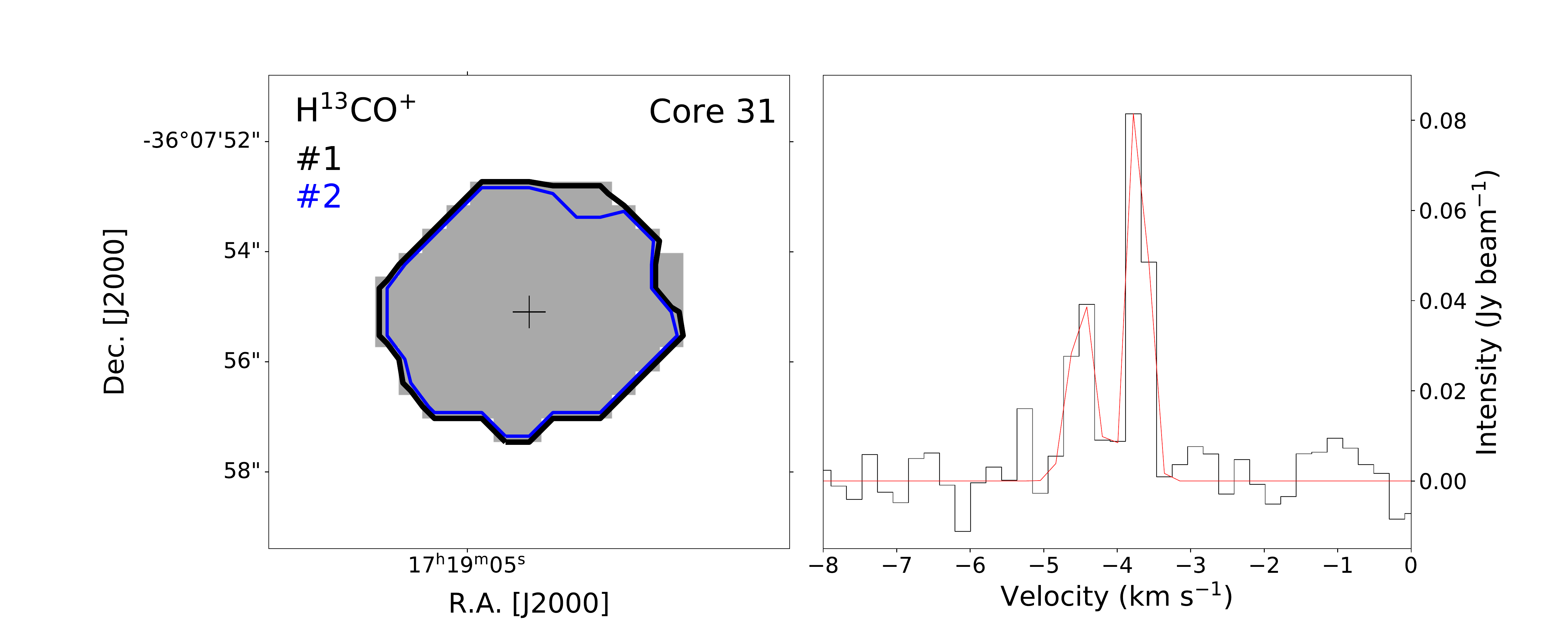}  \\
\includegraphics[clip,width=0.76\hsize]{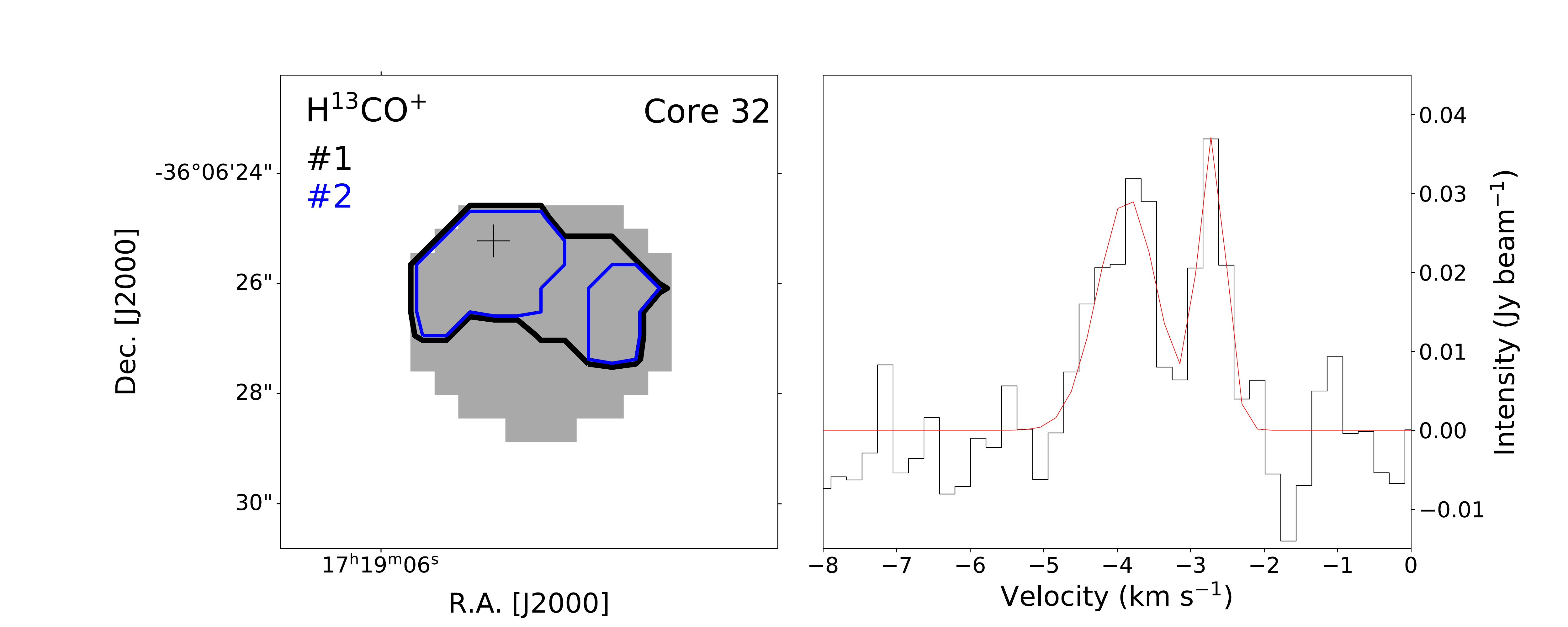}  \\
\includegraphics[clip,width=0.76\hsize]{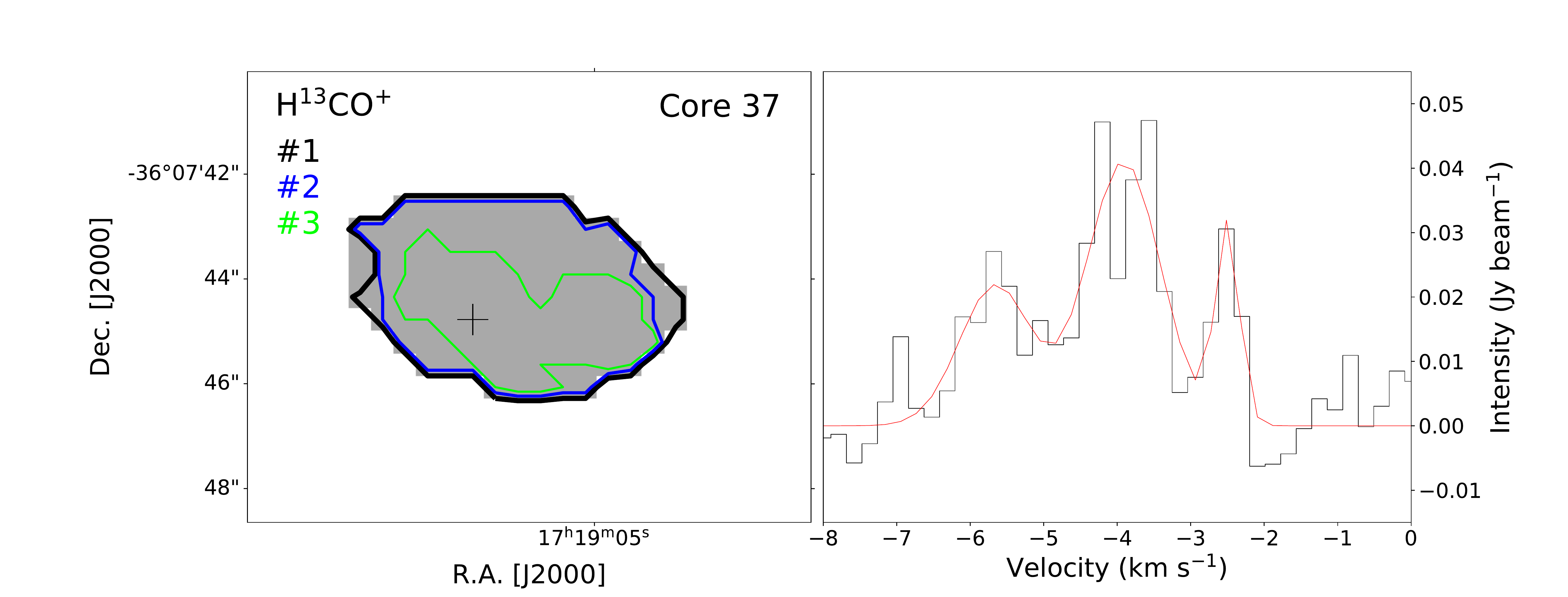} 
\end{tabular}

\newpage
\begin{center}
Presented here are individual plots for Figure~\ref{decomposed} 
which are available as the online material. 
\end{center}
\begin{tabular}{c}
\includegraphics[clip,width=0.76\hsize]{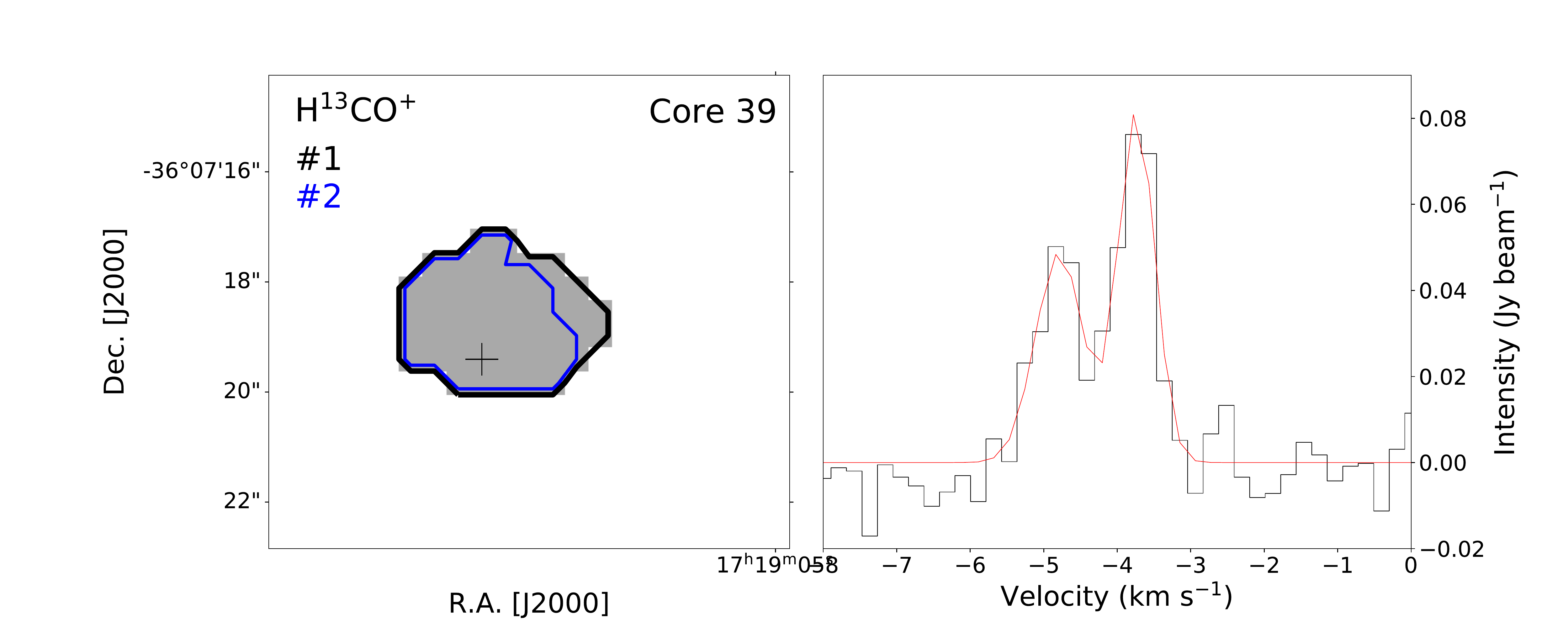}  \\
\includegraphics[clip,width=0.8\hsize]{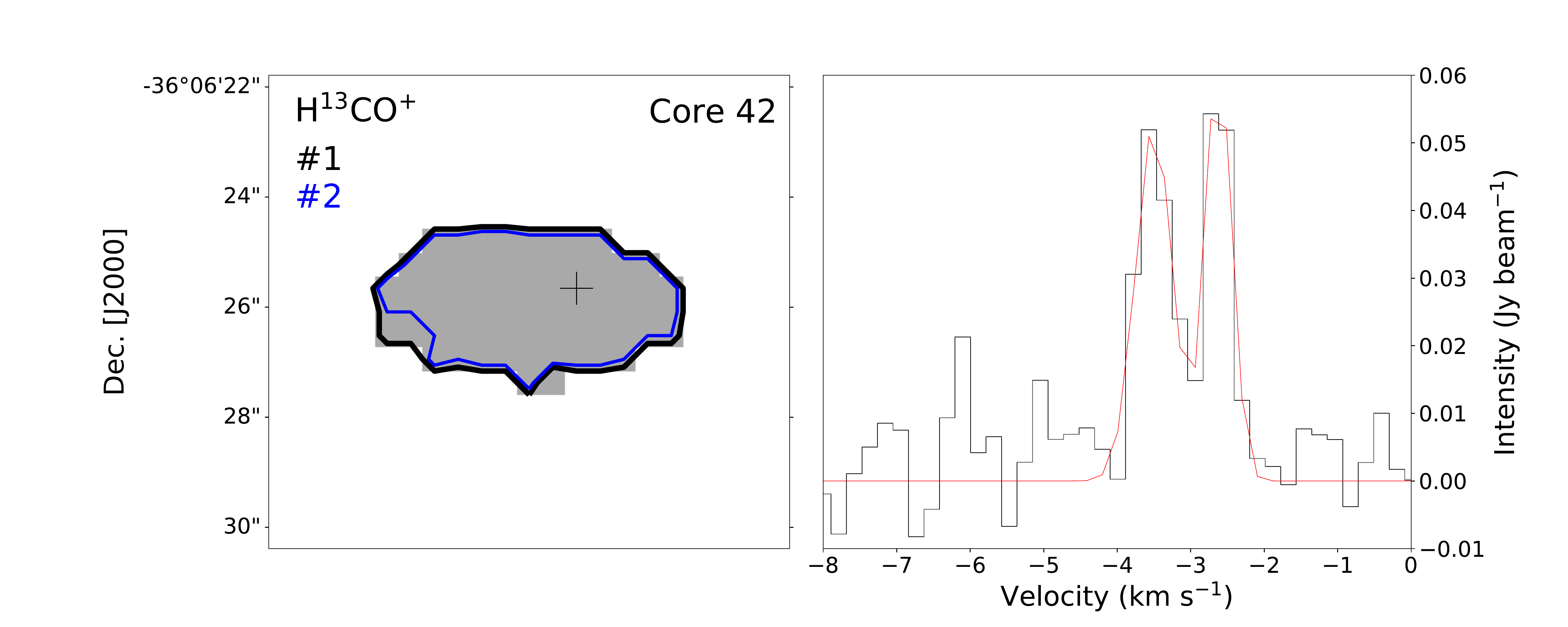}  \\
\includegraphics[clip,width=0.8\hsize]{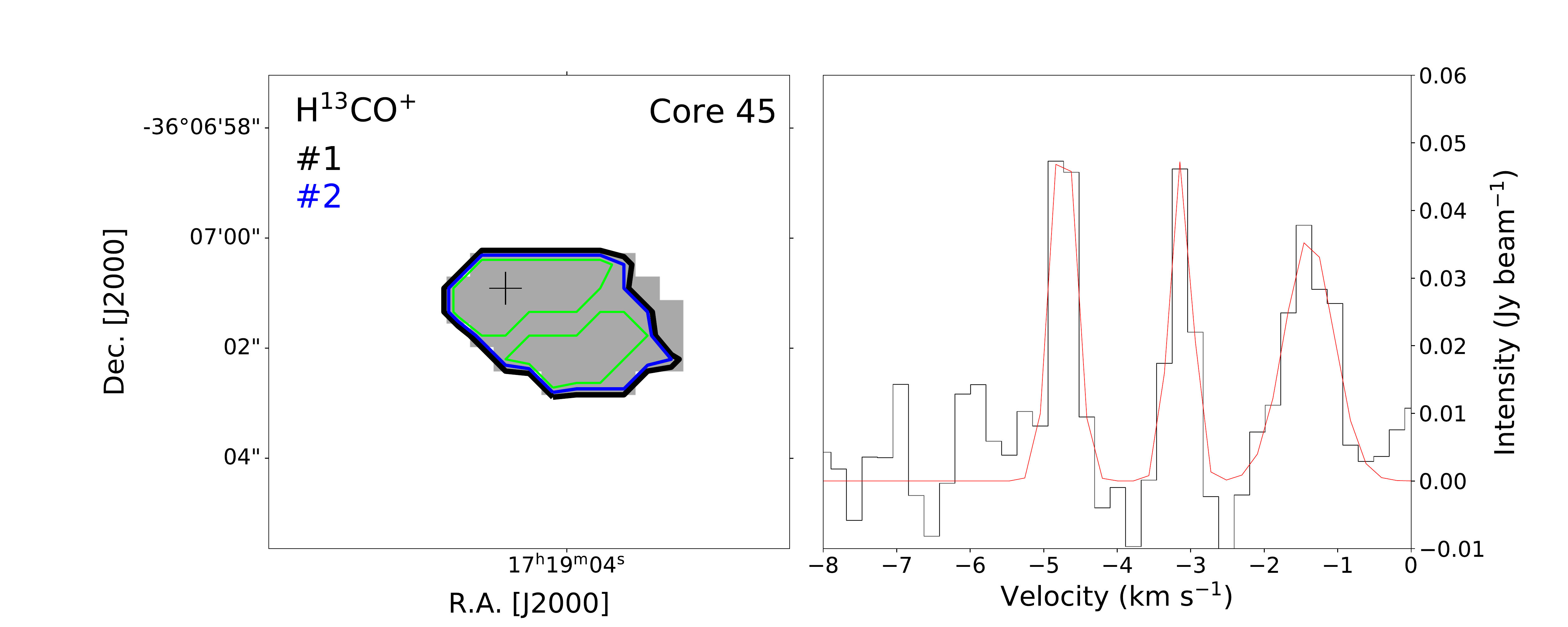} 
\end{tabular}

\end{document}